\documentclass[aps, prx, twocolumn, superscriptaddress, longbibliography]{revtex4-2}

\usepackage{graphicx}
\usepackage{amsmath}
\usepackage{amssymb}
\usepackage{float}
\usepackage{hyperref}
\hypersetup{
 pdftitle = {Long-time equilibration can determine transient thermality},
 pdfauthor = {K. Hovhannisyan, S. Nemati, C. Henkel, J. Anders},
 breaklinks=true,
 pdfnewwindow=true,
 colorlinks=true,
 linkcolor=red,
 citecolor=blue,
 filecolor=blue,
 urlcolor=blue
}
\usepackage[dvipsnames]{xcolor}
\usepackage{psfrag}
\usepackage{soul}
\usepackage{textcomp}
\usepackage{bbm}
\usepackage{bbold}
\usepackage{yfonts}
\usepackage[abs]{overpic}
\usepackage{pict2e}
\usepackage{pifont}
\usepackage{bbding}
\usepackage{tikz}
\usepackage{array}
\usepackage{multirow}
\usepackage{MnSymbol}
\usepackage[textfont=bf]{caption}
\usepackage{ragged2e}
\usepackage{subcaption}
\DeclareCaptionJustification{justified}{\justifying}

\newtheorem{theorem}{Theorem}
\newtheorem{corollary}{Corollary}[theorem]

\DeclareMathOperator{\tr}{Tr}
\DeclareMathOperator{\diag}{diag}

\DeclareMathOperator{\arctanh}{arctanh}
\DeclareMathOperator{\dist}{dist}

\newcommand{\comments}[1]{}
\newcommand{\bea}{\begin{eqnarray}}
\newcommand{\eea}{\end{eqnarray}}
\newcommand{\besa}{\begin{subequations}\begin{eqnarray}}
\newcommand{\eesa}{\end{eqnarray} \end{subequations}}
\newcommand{\beaa}{\begin{eqnarray}\begin{aligned}}
\newcommand{\eeaa}{\end{aligned}\end{eqnarray}}

\newcommand{\av}[1]{\left\langle #1 \right\rangle}

\newcommand{\id}{\mathbbm{I}}

\newcommand{\GGE}{\mathrm{GGE}}
\newcommand{\MF}{\mathrm{MF}}
\newcommand{\I}{\mathrm{int}}
\newcommand{\T}{\mathrm{T}}
\newcommand{\equi}{\mathrm{eq}}
\newcommand{\eff}{\mathrm{eff}}
\newcommand{\can}{\mathrm{can}}
\newcommand{\tot}{\mathrm{tot}}
\newcommand{\trel}{t_{\mathrm{eq}}}
\newcommand{\trev}{t_{\mathrm{rec}}}
\newcommand{\rhoseq}{\rho_s^{\mathrm{eq}}}

\newcommand{\cD}{\mathcal{D}}

\newcommand{\cE}{\mathcal{E}}
\newcommand{\cF}{\mathcal{F}}
\newcommand{\Dmins}{\mathcal{D}_s^{\min}}
\newcommand{\AT}{\mathrm{AT}}
\newcommand{\TT}{\tilde{t}}

\newcommand{\glocal}{g-local}
\newcommand{\Glocal}{G-local}

\begin{document}

\title{Long-time equilibration can determine transient thermality}

\author{Karen V. Hovhannisyan}
\email{karen.hovhannisyan@uni-potsdam.de}
\affiliation{University of Potsdam, Institute of Physics and Astronomy, Karl-Liebknecht-Str. 24-25, 14476 Potsdam, Germany}

\author{Somayyeh Nemati}
\affiliation{University of Potsdam, Institute of Physics and Astronomy, Karl-Liebknecht-Str. 24-25, 14476 Potsdam, Germany}

\author{Carsten Henkel}
\affiliation{University of Potsdam, Institute of Physics and Astronomy, Karl-Liebknecht-Str. 24-25, 14476 Potsdam, Germany}

\author{Janet Anders}
\email{janet@qipc.org}
\affiliation{University of Potsdam, Institute of Physics and Astronomy, Karl-Liebknecht-Str. 24-25, 14476 Potsdam, Germany}
\affiliation{Department of Physics and Astronomy, University of Exeter, Stocker Road, Exeter EX4 4QL, UK}

\begin{abstract}

When two initially thermal many-body systems start interacting strongly, their transient states quickly become non-Gibbsian, even if the systems eventually equilibrate. To see beyond this apparent lack of structure during the transient regime, we use a refined notion of thermality, which we call g-local. A system is g-locally thermal if the states of all its small subsystems are marginals of global thermal states. We numerically demonstrate for two harmonic lattices that whenever the total system equilibrates in the long run, each lattice remains g-locally thermal at all times, including the transient regime. This is true even when the lattices have long-range interactions within them. In all cases, we find that the equilibrium is described by the generalized Gibbs ensemble, with three-dimensional lattices requiring special treatment due to their extended set of conserved charges. We compare our findings with the well-known two-temperature model. While its standard form is not valid beyond weak coupling, we show that at strong coupling it can be partially salvaged by adopting the concept of a g-local temperature.

\end{abstract}

\maketitle

\section{Introduction}

Equilibration and thermalization in closed quantum many-body systems have received a lot of attention during the past two decades, leading to tremendous successes in understanding the conditions under which equilibration happens \cite{Gemmer_2004, Polkovnikov_2011, Gogolin_2016, Mori_2018} and the properties of the (sometimes thermal) equilibrium itself \cite{Gemmer_2004, Orus_2014, Gogolin_2016, Vidmar_2016, Trushechkin_2022, Alhambra_2022}. However, only two general ``expected behaviors'' are known about the \textit{transient} regime \footnote{By qualifying a behavior for a setup as ``expected'', we emphasize that it is proven to occur for the setup under certain restrictions, is known to sometimes occur also beyond those restrictions, but exceptions are also known. Such a state of affairs is common in statistical physics.}.
First, for a small subsystem weakly coupled to the rest of the large system one expects Markovianity of the dynamics \cite{Ford_1988, bp} \footnote{For a limited class of observables variables, a form of Markovianity can hold under more general conditions \cite{Strasberg_2023}.}. Second, when two well-separated relaxation timescales are present, some observables will typically show pre-thermalization \cite{Berges_2004, Gring_2012, Neyenhuis_2017, Tang_2018, Mori_2018}.
In this work, we propose a qualitatively new transient behavior for a generic physical setting
and then provide numerical evidence demonstrating that it indeed occurs in harmonic lattices.

The setting we consider---generic in mesoscopic and macroscopic physics---is that of two large quantum many-body systems, $X=A$ and $B$, of comparable size. Initially they do not interact and start uncorrelated, each in a \textit{global} Gibbs state 
\bea \label{eq:globalthermality}
	\tau(T_X, H_X) := \frac{1}{Z_X} \, e^{- H_X / T_X} .
\eea
Here $H_X$ is the Hamiltonian of $X$ and $T_X$ is a temperature in units $k_{\mathrm{B}} = 1$, with $Z_X:= \tr[e^{- H_X / T_X}]$ being the partition function. Because this (standard) definition focuses on each whole many-body system, we call it \textit{global thermality}. 

\begin{figure}[!t]
\centering
\includegraphics[width = 0.88\columnwidth]{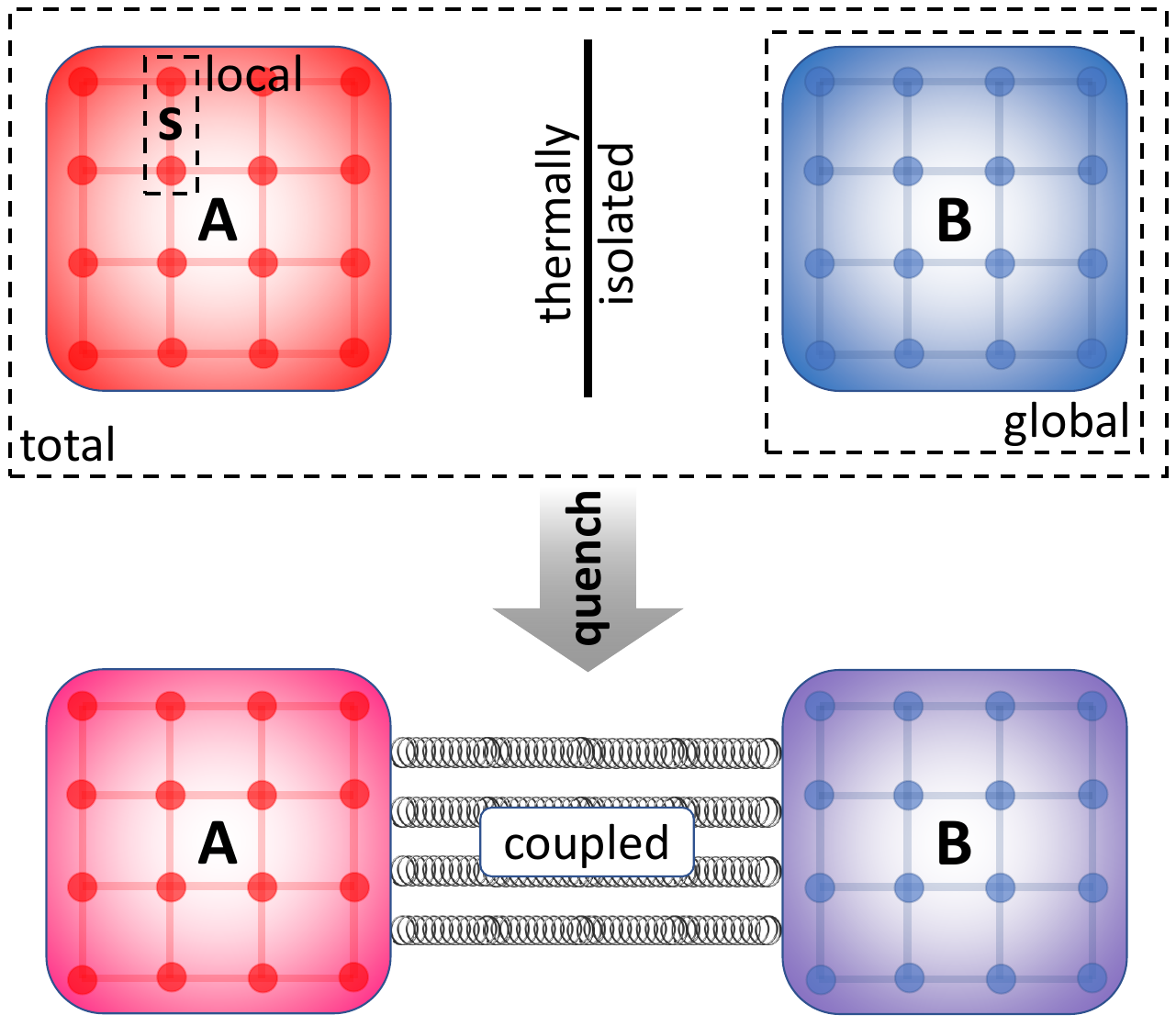}
\caption{(\textbf{General setup.})
Two quantum many-body systems, $X=A$ and $B$, start non-interacting and uncorrelated, each in a (global) Gibbs state Eq.~\eqref{eq:globalthermality}.
Then, in a sudden quench, interaction between them is switched on. As a result, neither $A$ nor $B$ will generally maintain global thermality during the joint post-quench unitary evolution. Here we establish in what sense a more local notion of thermality may be maintained during the evolution. Dashed boxes illustrate our scale terminology: ``total'' refers to $AB$-wide, ``global'' to either $A$ or $B$, and ``local'' pertains to small subsystems $s$ within $A$ or $B$. 
}
\label{fig:illustr}
\end{figure}

Then, in a sudden quench, coupling between the two systems is switched on, as depicted in Fig.~\ref{fig:illustr}. The total system state $\rho_{AB}(t)$ then evolves under the unitary evolution generated by the post-quench constant total Hamiltonian $H_\tot := H_A \otimes \id_B + \id_A \otimes H_B + H_\I$, where $H_\I$ is the interaction term and $\id_X$ is the identity operator on the Hilbert space of $X$.

The textbook expectation for weakly coupled macroscopic systems $A$ and $B$ is that the evolution progresses quasistatically and thus each of them retains global thermality [see Eq.~\eqref{eq:globalthermality}] at all times $t$, while gradual heat exchange brings the systems to a shared thermal equilibrium \cite{ll5}. In other words, the individual states of $A$ and $B$ obey $\rho_X (t) \approx \tau(T_X (t), H_X)$, with evolving temperatures $T_X(t)$ such that $T_A (t), T_B (t) \to T^\equi$. When the coupling is such that thermal gradients arise within $X$, still, the expectation would be that each small, localized portion of $X$ maintains a Gibbs state with respect to its local Hamiltonian \cite{ll5, Polder_1971, Cahill_2003, Volokitin_2007} at all times. Down to the mesoscopic scale, the assumption of instantaneous (local) thermality is the cornerstone of the two-temperature model (TTM) in solid-state physics \cite{Anisimov_1967, Anisimov_1974, Sanders_1977, Allen_1987}. It is designed to describe the joint dynamics of electrons and phonons in a solid after the electrons are suddenly heated up by radiation. Due to its simplicity, the TTM has been extensively employed for fitting the results of experiments and \textit{ab initio} calculations \cite{Sanders_1977, Allen_1987, Jiang_2005, Lin_2008, Wang_2012, Waldecker_2016, Pudell_2018, Herzog_2022}.

However, when the coupling between and within the two global systems is not weak, then neither the assumption of global thermality, nor that of local thermality, of each $A$ and $B$ is valid any longer \cite{Ferraro_2012, Kliesch_2014, Hernandez-Santana_2015, Intravaia_2016}.
Taking this general observation as a starting point, in this article we ask whether, and in what sense, these many-body systems may nevertheless keep appearing globally thermal when observed locally, i.e., on small subsystems.

To answer this, we begin by stating a new framework of thinking about thermality in Section~\ref{sec:glocal}, which subsumes the standard definition of thermality \eqref{eq:globalthermality}. It relies on both global and local properties of the system, and hence defines a new concept of thermality which we call ``{\glocal}.''
Its efficacy is demonstrated on harmonic lattices, a realistic yet efficiently simulable system \cite{Ford_1965, Mermin76, ll5, Ford_1988, bp}, which we introduce in Section~\ref{sec:setup}. By numerically solving their dynamics, we establish in Section~\ref{sec:glocality} how well {\glocal} thermality captures the instantaneous states of the co-evolving systems. 
In Section~\ref{sec:equil}, we look at the process of equilibration and discuss the subtleties of constructing the generalized Gibbs ensemble (GGE) describing it.
We close with a brief discussion of implications for the validity of the TTM ansatz in Section~\ref{sec:TTM}, before concluding in Section~\ref{sec:discussion}. 

Our main result is that \textit{if} local observables of the total system $AB$ equilibrate for long times, then each $A$ and $B$ maintain {\glocal} thermality to a very good approximation at all times, including the transient regime. Moreover, this behavior is valid at all coupling strengths, including very strong coupling. This all-time validity of {\glocal} thermality is surprising because, in general, the dynamics during the transient regime is thought to be structureless. The result thus fleshes out a novel ``expected behavior'' for the process of joint equilibration of two large systems.

\section{{\glocal} thermality}
\label{sec:glocal}

Consider a state $\rho_X$ of a many-body system $X$ and a small local subsystem $s \subset X$; see Fig.~\ref{fig:illustr}. We ask whether a temperature $T$ exists such that the reduced state $\rho_s = \tr_{X \backslash s}[\rho_X]$ of $s$ obeys
\bea \label{def:g-local}
	\rho_s \overset{?}{=} \tr_{X \backslash s} [\tau(T, H_X)],
\eea
where the partial trace is taken over all of $X$ except $s$. If this condition is obeyed, then we say that $X$ is ``{\glocal}ly thermal at $s$.'' The term ``{\glocal}'' is to emphasize that, while $\rho_s$ is a \textit{local} quantity, it contains information about the \textit{global} $\tau(T, H_X)$ due to non-negligible interactions within $X$. Furthermore if $X$ is {\glocal}ly thermal at each small subsystem, then we call $X$ ``{\glocal}ly thermal.'' If in addition $T$ is the same for all of them, then we call $X$ ``uniformly {\glocal}ly thermal.'' Otherwise, when $T$ varies depending on the subsystem, we say that $X$ is ``{\glocal}ly thermal with a gradient.''

Note that condition \eqref{def:g-local} is not to be confused with subsystem $s$ being in a Gibbs state $\tau(T,H_s)$ at $T$ with respect to its local (bare) Hamiltonian $H_s$. 
Indeed, it is well-known that $\rho_s$ can differ significantly from $\tau(T, H_s)$ \cite{Onsager_1933, Kirkwood_1935, Haake_1985, Ferraro_2012, Kliesch_2014, Hernandez-Santana_2015, Seifert_2016, Jarzynski_2017, Miller_2018Rev, Alhambra_2022}. Instead, partially reduced states of global Gibbs states 
\bea \label{eq:thermala}
    \tau_s^\MF(T) := \tr_{X \backslash s} [\tau(T, H_X)]
\eea
are known as ``mean force (Gibbs) states'' \cite{Cresser_2021, Trushechkin_2021, Latune_2022, Trushechkin_2022}.
With this definition, the condition of ``{\glocal} thermality of $X$'' can be compactly expressed as
\bea \label{eq:exact}
    \rho_s \, \overset{!}{=} \, \tau_s^\MF(T), \quad \forall s \subset X,
\eea
where the $s$ are small subsystems.

However, in most realistic scenarios one cannot expect the equality \eqref{eq:exact} to be exact. Thus, it is sensible to introduce an effective {\glocal} temperature $T^\eff_s$ for each subsystem $s \subset X$ as that of the mean force state $\tau_s^\MF(T)$ that is closest to $\rho_s$. Namely,
\bea \label{teffdef}
    T^\eff_s := \arg \min_T \cD\left[\rho_s, \; \tau_s^\MF(T) \right],
\eea
where as a measure of distance $\cD$ between the two states, we chose the Bures metric \cite{mikeike} (see Appendix~\ref{app:dist} for the definition). The distance
\bea \label{buresdef}
    \Dmins := \min_T \cD\left[ \rho_s, \; \tau_s^\MF(T) \right]
\eea
then measures to what extent $\rho_s$ deviates from the optimal mean-force Gibbs state.
In what follows, we will use the dual quantity, the \textit{fidelity} \cite{mikeike},
\bea \label{fideldef}
    \cF^{\max}_s := \big[1 - (\Dmins)^2 / 2\big]^2 \leq 1,
\eea
and call this the degree of {\glocal} thermality of $X$ at $s$. The fidelity is $1$ \textit{iff} the two states $\rho_s$ and $\tau_s^\MF$ are equal, in which case $T_s^\eff$ turns into a proper {\glocal} temperature for $s$. Therefore, the higher the $\cF_s$, the closer the local system $s$ is to having a well-defined {\glocal} temperature; see Eq.~\eqref{teffdef}.

The pair $(T_s^\eff, \cF^{\max}_s)$ thus fully characterizes the {\glocal} thermality of $X$ at subsystem $s$. If the $T_s^\eff$ for essentially all small $s \subset X$ are approximately equal to each other, and all $\cF^{\max}_s$'s are close to $1$ (within a chosen error \footnote{As a guideline, target values $\cF \geq 0.99$ for the fidelity ($\cD \leq 0.1$) are considered high in current quantum technologies (see, e.g., Refs.~\cite{Bradley_2019, Zhou_2020, Rudolph_2022})}), then $\rho_X$ (or $X$ itself) is {\glocal}ly thermal, with uniform temperature $T^\eff_X$. In section \ref{sec:glocality}, we will use $T_s^\eff$ and $ \cF^{\max}_s$ to assess the {\glocal} thermality of each of the two global systems, $A$ and $B$, of comparable size $N$, during their joint evolution.

\medskip

Unmistakably, our framework is inspired by the equivalence of ensembles \cite{Simon, Muller_2015, Brandao_2015, Tasaki_2018, Kuwahara_2020} and canonical typicality \cite{Goldstein_2006, Popescu_2006, Gogolin_2016}. The difference is in how temperature is defined. There, the effective temperature $T_X^{\eff, \can}$ is determined by equating the mean energies, i.e.,
\bea \label{Teff_can}
	\tr\big[\tau\big(T_X^{\eff, \can}, H_X\big) H_X\big] = \tr[\rho_X H_X],
\eea
and it is shown that Eq.~\eqref{def:g-local} is satisfied for $T=T_X^{\eff, \can}$ under certain conditions on $H_X$ and $\rho_X$. 
This approach is thus energy-centric and global: $T_X^{\eff, \can}$ is the same for all subsystems. In contrast, our framework is state-centric and local: it directly accesses the marginal state of a subsystem $s$ and defines $T^\eff_s$ as the solution of the optimization problem \eqref{buresdef}. The ability to define a local degree of thermality and an associated temperature at each subsystem allows our framework to accommodate systems with a temperature gradient (see Appendix~\ref{app:EEdyn} for an example), which is beyond the reach of the typicality-based approaches.
Note that, when $H_X$ is a sum of local terms and the system is {\glocal}ly thermal with a uniform {\glocal} temperature $T_X^\eff$, then the two temperatures coincide: $T^{\eff, \can}_X = T^\eff_X$ (see Appendix~\ref{app:local_meas}).

Lastly, when the size of the system $X$ is finite, then Eq.~\eqref{def:g-local} will hold for a system in a canonically typical state only approximately, with the correction going to zero as $N_s/N_X \to 0$, where $N_s$ and $N_X$ are the numbers of sites in $s$ and $X$, respectively. Similarly, in our framework, we expect $\Dmins$ to also have a positive contribution stemming from the small parameter $N_s/N_X$ in realistic scenarios. This finite-size contribution will likely be a highly complex function of $N_s/N_X$ \cite{Popescu_2006, Farrelly_2017}, and the line between ``small'' and ``big'' subsystems will be drawn by this system- and situation-dependent contribution and one's error tolerance. Importantly, the finite-size effect will in general not be the only factor contributing to $\Dmins$.

\begin{figure}[!t]
\centering
\includegraphics[width = 0.88\columnwidth]{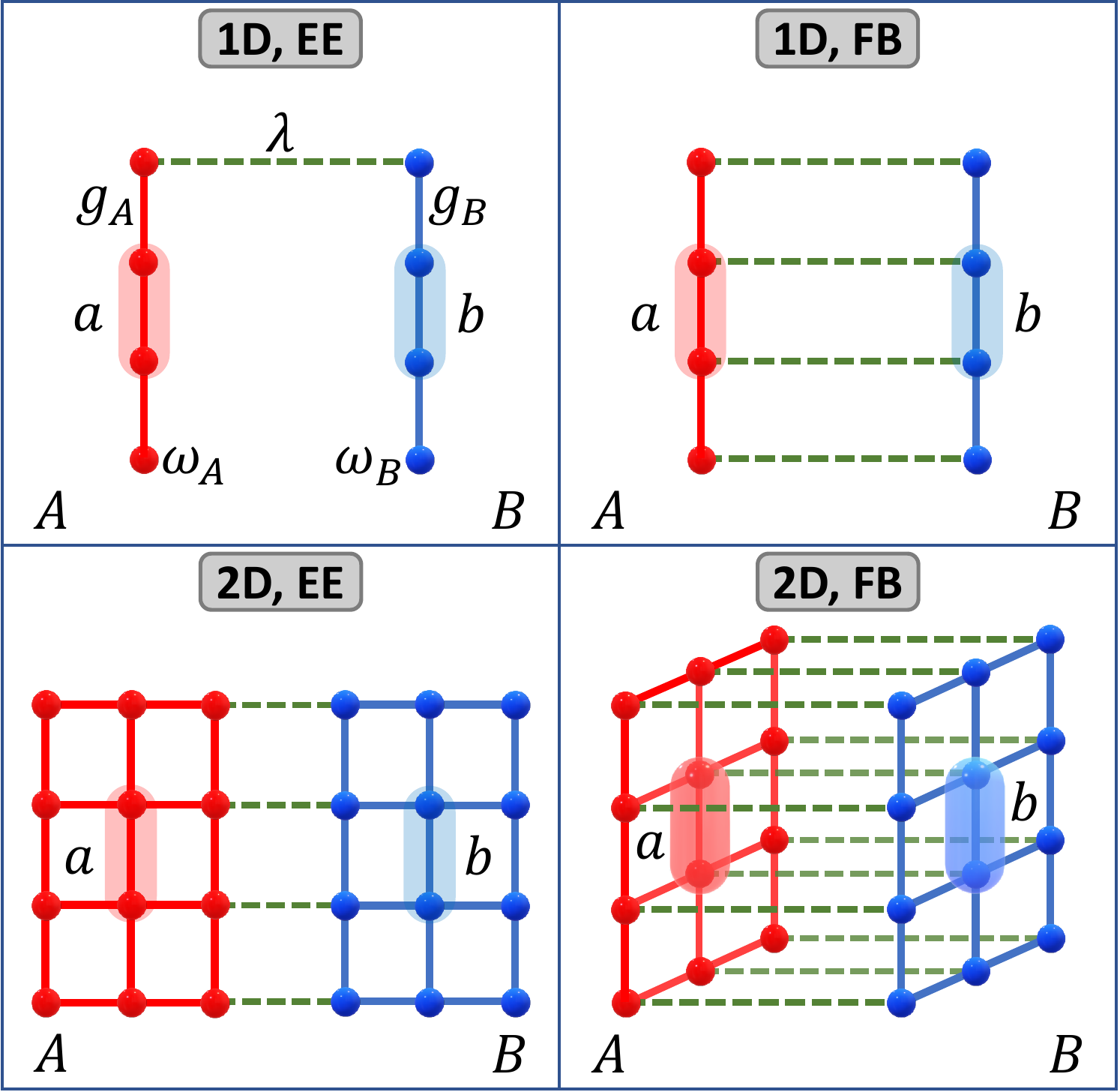}
\caption{
(\textbf{Model schematic.}) Systems $A$ (red) and $B$ (blue) are here modeled as 1D (\textbf{top row}) or 2D (\textbf{bottom row}) harmonic lattices. Each site (circles) denotes a local oscillator of frequency $\omega_X$ which is coupled to its neighbors with strength $g_X$ (solid lines), cf. Eq.~\eqref{hamilton}. While only nearest-neighbor interactions are depicted, our results apply also to long-range interacting systems.
Inter-system coupling (dashed green lines) with strength $\lambda$ occurs either only at the system edges---as shown in the \textbf{left column} (edge--edge (EE) coupling), or at all sites---as shown on the \textbf{right column} (full-body (FB) coupling), see Eq.~\eqref{hint}. 
In all panels, example subsystems $a$ and $b$ are depicted, at which {\glocal} thermality of each, $A$ and $B$, will be assessed in Sec.~\ref{sec:glocality}.
}
\label{fig:couplings}
\end{figure}

\section{Setup and model} 
\label{sec:setup}

As mentioned in the introduction, our setup consists of two large many-body systems, $A$ and $B$, co-evolving after an interaction between them is switched on. To be able to solve the dynamics of the total system $AB$ and demonstrate the occurrence (or absence) of {\glocal} thermality of each, $A$ and $B$, we chose harmonic lattices. Despite their simplicity, these systems are routinely used to approximate various physical systems
\cite{Ford_1965, Mermin76, ll5, Ford_1988, bp}. At the same time, the dynamics of the Gaussian states in these systems admit a numerically efficient phase-space representation \cite{Adesso_2007, Anders_2008, Weedbrook_2012}, allowing us to directly simulate few-hundred-particle lattices.

Each global system is a 1D or 2D translation-invariant open-ended lattice, see Fig.~\ref{fig:couplings}, with Hamiltonian
\bea \label{hamilton}
	H_X = \sum_{\nu} \bigg[\frac{\omega_X^2 q_{X, \nu}^2}{2} + \frac{p_{X, \nu}^2}{2} \bigg]
	+ \sum_{\nu, \nu'} G_X^{\nu, \nu'} q_{X, \nu} \, q_{X, \nu'}, ~~~
\eea
where $\nu$ enumerates the sites in lattice $X$, $\omega_X$ is the on-site frequency of each site, and all masses are set to $1$. The intra-system coupling function, $G_X^{\nu, \nu'}$, depends only on the distance between the sites $\nu$ and $\nu'$. Our numerical samples below explore lattices with coupling functions of the form
\bea \label{decayrate}
	G_X^{\nu, \nu'} = \frac{g_X}{\dist(\nu, \nu')^\alpha},
\eea
where $\dist(\nu, \nu')$ is the Manhattan distance between the sites $\nu$ and $\nu'$, and $\alpha > 0$ quantifies the \textit{range} of interactions. Nearest-neighbor interactions correspond to $\alpha = \infty$ (and couple only sites with $\dist(\nu, \nu') = 1$).

We recall that $A$ and $B$ are large and of comparable size. Therefore, for simplicity of presentation, we choose the lattices $A$ and $B$ to have the same size and shape, with $N \gg 1$ denoting the number of sites in each of them.
The opposite limit, where one of the systems is much smaller than the other, say, $N_A \ll N_B$, is well-understood in harmonic systems. $A$ then simply thermalizes with $B$, in the sense that its state tends to $\tr_{B}[\tau(T_B, H_{AB})]$ (save for finite-size effects) \cite{Tegmark_1994, Subasi_2012}.

For the interaction term between $A$ and $B$, $H_\I$, we consider two types of coupling: edge--edge (EE) and full-body (FB), shown in Fig.~\ref{fig:couplings} for 1D and 2D lattices. For example, the FB interaction has the form
\bea \label{hint}
    H_\I^{\mathrm{(FB)}} = \lambda \sum_{\nu} q_{A, \nu} \, q_{B, \nu},
\eea
where $\lambda$ is the inter-system coupling strength, and $\nu$ runs over \textit{all} corresponding sites in $A$ and $B$; see the right column of Fig.~\ref{fig:couplings}.
Given the form of Eqs.~\eqref{hamilton} and \eqref{hint}, the natural dimensionless coupling constants are $g_X/\omega_X^2$ and $\lambda/(\omega_A \omega_B)$.

\begin{figure}[!t]
\centering
\includegraphics[width=0.99\columnwidth]{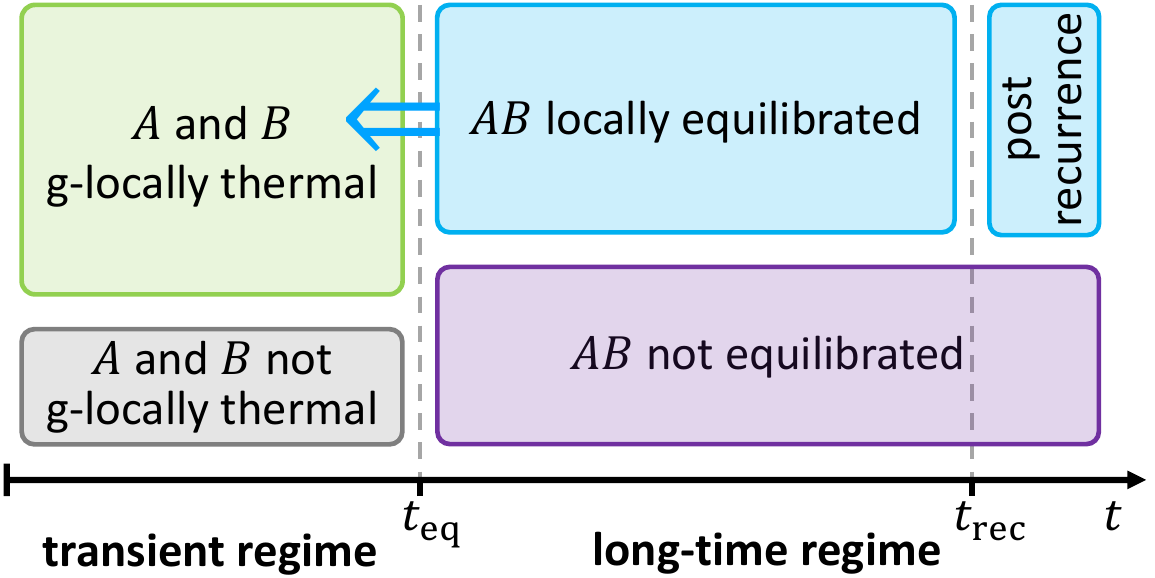}
\caption{
(\textbf{Illustration of main result.})
Our main result is that \textbf{if} local equilibration occurs at long times, i.e., $\rho_s(t) \approx \rho_s^\equi$ for all small subsystems $s \subset AB$ for $t \in [\trel, \trev]$ (see Sec.~\ref{sec:equil}), 
\textbf{then} $A$ and $B$ are {\glocal}ly thermal at any time $t$, also in the transient regime (see Sec.~\ref{sec:glocality}). 
Namely, $\rho_s(t) \approx \tau_s^\MF \left(T^\eff_s(t)\right)$, where $T^\eff_s(t)$ is a time-dependent {\glocal} temperature of subsystem $s$ (see Sec.~\ref{sec:glocal}). 
Remarkably, this result characterizes the transient regime and links it with the long-time equilibration behavior of the system.
}
\label{fig:sketch}
\end{figure}

For the initial state, we take the uncorrelated state
\bea \label{rhonin}
	\rho_{AB}(0) = \tau (T_A, H_A) \otimes \tau (T_B, H_B),
\eea
and the evolution of the joint system $AB$ is generated by the total Hamiltonian $H_\tot$.
The Gaussian theory that underpins the simulation of harmonic systems has been reviewed, e.g., in Refs.~\cite{Adesso_2007, Anders_2008, Weedbrook_2012}. We give a brief account of the main quantities and formulas used in our simulations in Appendix~\ref{app:Gaussian}.
Using these methods, we numerically solve the dynamics of [1D,EE], [1D,FB], [2D,EE] and [2D,FB] lattices for a representative selection of the full range of parameter values for which the spectrum of $H_\tot$ is bounded from below \footnote{For fixed $\omega_X$'s and $\alpha$, the requirement that $H_\tot$ must be bounded from below sets an upper bound on $|g_X|$ and $|\lambda|$.}.

Our direct simulation of the dynamics of the total system $AB$ gives us access to $\rho_A(t)$ and $\rho_B(t)$, which allows us to analyze the {\glocal} thermality of $A$ and $B$ at all times during their joint post-quench evolution.

\section{All-time {\glocal} thermality}
\label{sec:glocality}

We have performed a large number of numerical experiments spanning the full parameter range, and established the following: {\Glocal} thermality of $A$ and $B$ is guaranteed \textit{at all times, including transient times}, whenever \textit{all local observables of $AB$ equilibrate dynamically at long times}; see Sec.~\ref{sec:equil} for further details on this requirement. This behavior occurs for all intra-lattice coupling strengths $g_X$ and interaction ranges $\alpha$ and inter-lattice couplings $\lambda$. This is the first main result of the paper. An illustration of this relation between long-time and transient behavior is shown in Fig.~\ref{fig:sketch}. A detailed account on how we perform the numerical proof, as well as the numerical evidence itself, can be found in Appendix~\ref{app:numerics}.

\begin{figure}[!t]
\centering
\includegraphics[trim=0.0cm 0.3cm 0cm 0cm,clip, width=\columnwidth]{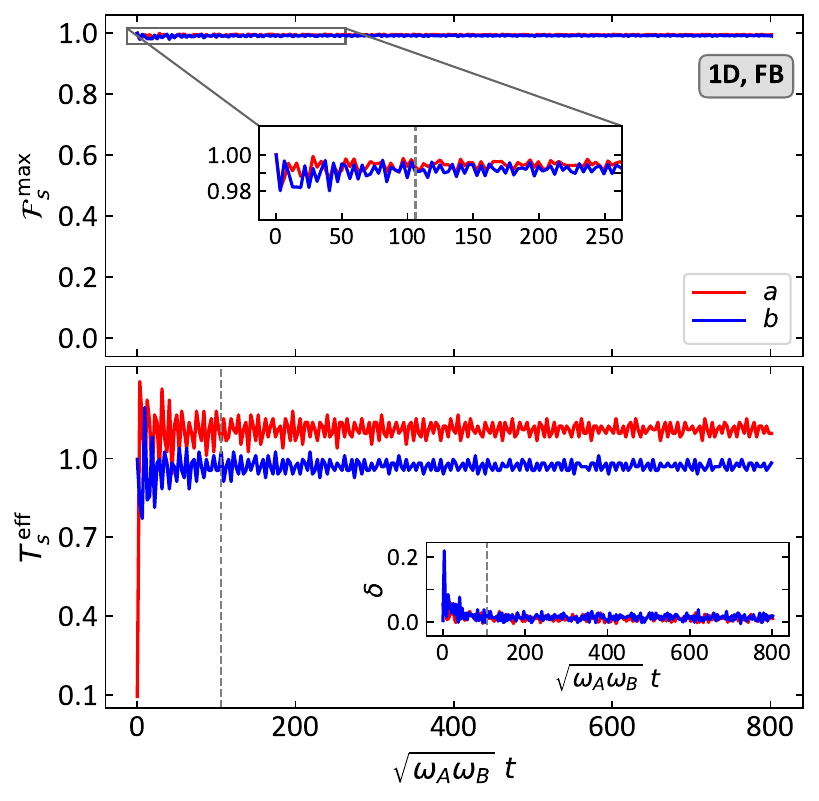}
\caption{
(\textbf{{\Glocal} thermality of $A$ at $a$ and $B$ at $b$.}) 
\textbf{Top panel:} the fidelity $\cF^{\max}_{s}$~\eqref{fideldef}, which measures the degree of {\glocal} thermality of $A$ at $s=a$ (red) and of $B$ at $s=b$ (blue), as a function of time $t$. The inset zooms into the fidelity at early times.
The \textbf{bottom panel} shows the corresponding effective {\glocal} temperatures $T_{s}^\eff$ defined in Eq.~\eqref{teffdef}. The inset shows the normalized difference $\delta = (T_X^{\eff, \can} - T_X^\eff)/T_X^\eff$ between the {\glocal} and effective canonical~\eqref{Teff_can} temperatures.
The fidelities are close to $1$, indicating that both $A$ and $B$ are {\glocal}ly thermal at $a$ and $b$, respectively, with very good precision \textit{at all times} during the evolution.
Note that subsystems $a$ and $b$ settle to slightly different {\glocal} temperatures.
This plot is for $a$ and $b$ each consisting of two consecutive sites situated at the centers of the 1D chains $A$ and $B$, respectively. The chains are $N_A = N_B = 200$ long and interact through full-body (FB) coupling. Each chain features long-range interactions, with the decay rate $\alpha = 1/2$ [cf.~Eq.~\eqref{decayrate}]. The rest of the Hamiltonian parameters are $\omega_A = 2$ and $g_A / \omega_A^2 = 0.2$ for $A$, $\omega_B = 1.5$ and $g_B / \omega_B^2 = 0.3$ for $B$, and the inter-chain coupling is $\lambda/(\omega_A \omega_B) = 0.5$. The initial temperatures are $T_A = 0.1$ and $T_B = 1$.
The vertical dashed lines in both panels indicate the instance at which Fig.~\ref{fig:gloc_site_1DFB} is plotted.
}
\label{fig:gloc_time_1DFB}
\end{figure}

\begin{figure}[!t]
\centering
\includegraphics[trim=0.0cm 0.3cm 0cm 0cm,clip, width=\columnwidth]{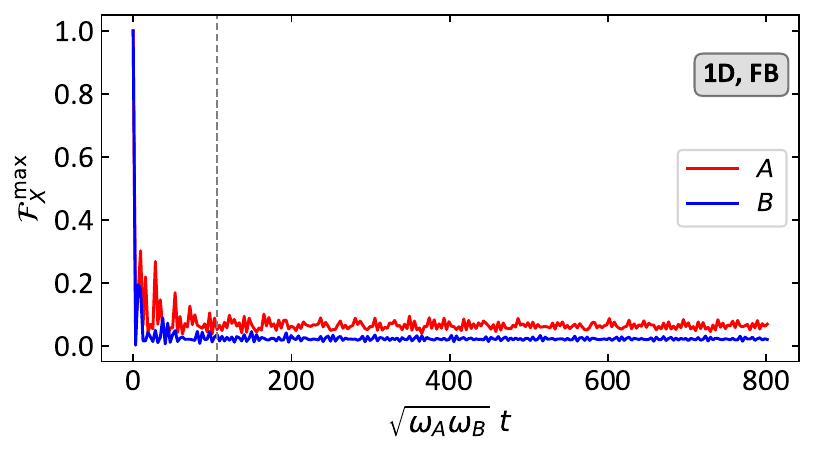}
\includegraphics[trim=0.0cm 0.3cm 0cm 0cm,clip, width=\columnwidth]{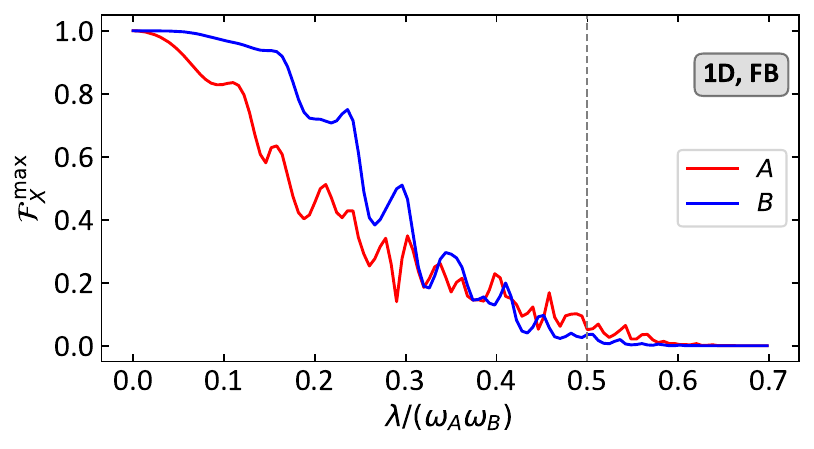}
\caption{
(\textbf{Global non-Gibbsianity of $A$ and $B$.}) 
\textbf{Top panel}: Fidelity $\cF^{\max}_X$ of the state $\rho_X (t)$ of many-body system $X = A, B$ with the closest global Gibbs state \eqref{eq:globalthermality} as a function of time $t$. Starting from a global Gibbs state for which the fidelity is $1$, the state of $X$ deviates increasingly from Gibbs form with increasing time of interaction between $A$ and $B$. The inter-chain coupling for this panel is $\lambda / (\omega_A \omega_B) = 0.5$, as in Fig.~\ref{fig:gloc_time_1DFB}. 
\textbf{Bottom panel}: Fidelity $\cF^{\max}_X$ of the state $\rho_X(t_0)$ at a fixed moment of time $t_0$ as a function of the coupling strength $\lambda$, for $\lambda \in [0, \lambda_{\max}]$, where $\lambda_{\max} \approx 0.695 \omega_A \omega_B$ is the largest value allowed for $|\lambda|$ in this configuration. The time $\sqrt{\omega_A \omega_B} \, t_0 = 106$ (the vertical dashed line in the top panel) is chosen so as to allow $AB$ to evolve considerably away from its initial state. The dashed line on this panel shows the coupling strength at which the top panel is plotted. Both panels are for the same [1D, FB] configuration, with all other parameters being the same as in Fig.~\ref{fig:gloc_time_1DFB}.
}
\label{fig:global_atherm_1DFB}
\end{figure}

An immediate practical consequence of this result is that, if an experimenter monitoring a small region of the system notices that {\glocal} thermality is violated at that location, then they can predict with certainty that the system will \textit{not} ever equilibrate as a whole.

An example illustration of the above general result is given by Fig.~\ref{fig:gloc_time_1DFB}. It shows the fidelities and effective temperatures for the case where $A$ and $B$ are both open-end 1D chains of $200$ sites with long-range interactions within them ($\alpha = 0.5$), which are coupled via a full-body (FB) interaction Hamiltonian; see Fig.~\ref{fig:couplings}. The plots in Fig.~\ref{fig:gloc_time_1DFB} are for subsystem $a$ consisting of two consecutive sites in the middle of chain $A$, and similarly for $b$ in $B$.
The top panel shows the degree of {\glocal} thermality of $A$ at $a$ (red), and of $B$ at $b$ (blue), as defined in Eq.~\eqref{fideldef}. As one can see, they are close to $1$ at all times, demonstrating the {\glocal} thermality of $A$ at $a$ and $B$ at $b$. The corresponding time-evolving effective {\glocal} temperatures $T_{a}^\eff$ and $T_{b}^\eff$ of the small subsystems $a$ (red) and $b$ (blue) are shown in the bottom panel. These {\glocal} temperatures slowly converge in time, while oscillating about each other. The apparent symmetric nature of these oscillations is due to the fact that the interaction energy remains small for the chosen set of parameters (see Sec.~\ref{sec:TTM} with Fig.~\ref{fig:heatflow} and the discussion in Appendix~\ref{app:local_meas}).

To appreciate the nontriviality of the high values of the fidelity in Fig.~\ref{fig:gloc_time_1DFB}, note that $\lambda = 0.5 \, \omega_A \omega_B$ corresponds to quite strong coupling. Indeed, it is close to the maximal coupling strength ($\lambda_{\max} \approx 0.695 \, \omega_A \omega_B$) consistent with the requirement that $H_\tot$ must be bounded from below, with all the other parameters fixed. Moreover, $\lambda / g_A = 1.875$ and $\lambda / g_B \approx 2.222$, which means that the coupling strongly perturbs all the nodes of both $A$ and $B$. Nonetheless, both $A$ and $B$ maintain a high degree of {\glocal} thermality ($\geq 0.98$) at all times. For comparison, for the not-much-larger $\lambda = 0.65 \, \omega_A \omega_B$, $\cF_a^{\max}$ and $\cF_b^{\max}$ get as low as, respectively, $0.93$ and $0.884$ during the evolution.

Moreover, the all-time high degree of {\glocal} thermality of $A$ and $B$ in Fig.~\ref{fig:gloc_time_1DFB} is in stark contrast with the quick loss of \textit{global} thermality by them, especially at higher coupling strengths. Similarly to Eqs.~\eqref{buresdef} and \eqref{fideldef}, we quantify the degree of \textit{global} thermality of system $X$ as $\cF_X^{\max}(t) := \max_T \mathcal{F}[\rho_X(t), \tau(T, H_X)]$. Fig.~\ref{fig:global_atherm_1DFB} shows the result for the same [1D, FB] system as in Fig.~\ref{fig:gloc_time_1DFB}. One sees that the degree of global thermality in Fig.~\ref{fig:global_atherm_1DFB} quickly drops from 1 to $\lesssim 0.05$, which is a clear indication that the global state is not Gibbsian \footnote{Note that the stabilisation of $\cF^{\max}_X$ does not imply that the states $\rho_X(t)$ themselves stabilize---just their distance from the set of Gibbs states does.}.

Lastly, we find that for most parameter choices for which the total system $AB$ \textit{does not} locally equilibrate at long times (purple box in Fig.~\ref{fig:sketch}), the global systems $A$ and $B$ do not develop stable {\glocal} thermality. However, there do exist parameter values for which $AB$ does not equilibrate, but $A$ and $B$ do still maintain {\glocal} thermality at all times.

\begin{figure}[t]
\centering
\includegraphics[trim=0.0cm 0.3cm 0cm 0cm,clip, width=\columnwidth]{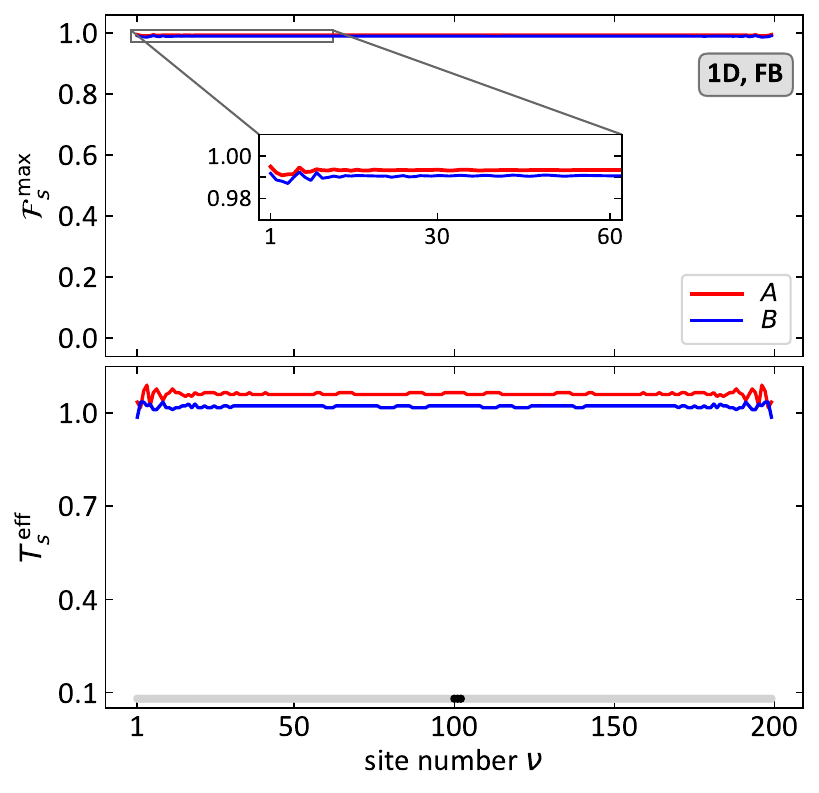}
\caption{
(\textbf{{\Glocal} thermality with site number $\nu$.})
The degree of {\glocal} thermality (\textbf{top panel}) $\cF^{\max}_s$~\eqref{fideldef} of $A$ at two-site subsystem $s=a$ (red), and of $B$ at $s=b$ (blue), respectively, and (\textbf{bottom panel}) the corresponding effective temperatures $T_s^\eff$~\eqref{teffdef} as a function of site number $\nu$. Here $\nu = 1, ..., N - 1$ labels \textit{all} subsystems consisting of two neighboring sites in $A$ and $B$. 
The fidelities are very close to 1 for all $\nu$. Furthermore, all subsystems $s$ located away from the edges, i.e., those ``inside the bulk'', give essentially the same {\glocal} temperature $T_{s}^\eff$. Together these two plots show that both $A$ and $B$ are uniformly {\glocal}ly thermal at \textit{all} two-site subsystems. 
The ``slider'' (black dot) at the bottom indicates the position of subsystem $s$ in $X$ at which Fig.~\ref{fig:gloc_time_1DFB} is plotted.
We highlight that this is just one representative snapshot of the {\glocal} properties of $A$ and $B$ at any particular moment in time; here $\sqrt{\omega_A \omega_B} \, t_0 = 106$. 
We observe qualitatively the same plots for \textit{any small subsystem size}, see Fig.~\ref{fig:gloc_subsize_1DFB}, and at other time points (with varying temperatures), see Fig.~\ref{fig:gloc_time_1DFB}.
All other parameters are as in Figs.~\ref{fig:global_atherm_1DFB} and~\ref{fig:gloc_time_1DFB}.
}
\label{fig:gloc_site_1DFB}
\end{figure}

Now moving on from the specific two-site subsystems located in the respective centers of the chains, the plots in Fig.~\ref{fig:gloc_site_1DFB} show that 1D chains $A$ and $B$ are {\glocal}ly thermal with respect to \textit{all} two-site subsystems $\nu$ along the chains. Moreover, ``inside the bulk'', i.e., away from the edges inwards, all two-site subsystems share the same temperature. Both $A$ and $B$ are thus uniformly {\glocal}ly thermal.

\begin{figure}[t]
\centering
\includegraphics[trim=0.0cm 0.3cm 0cm 0cm,clip, width=\columnwidth]{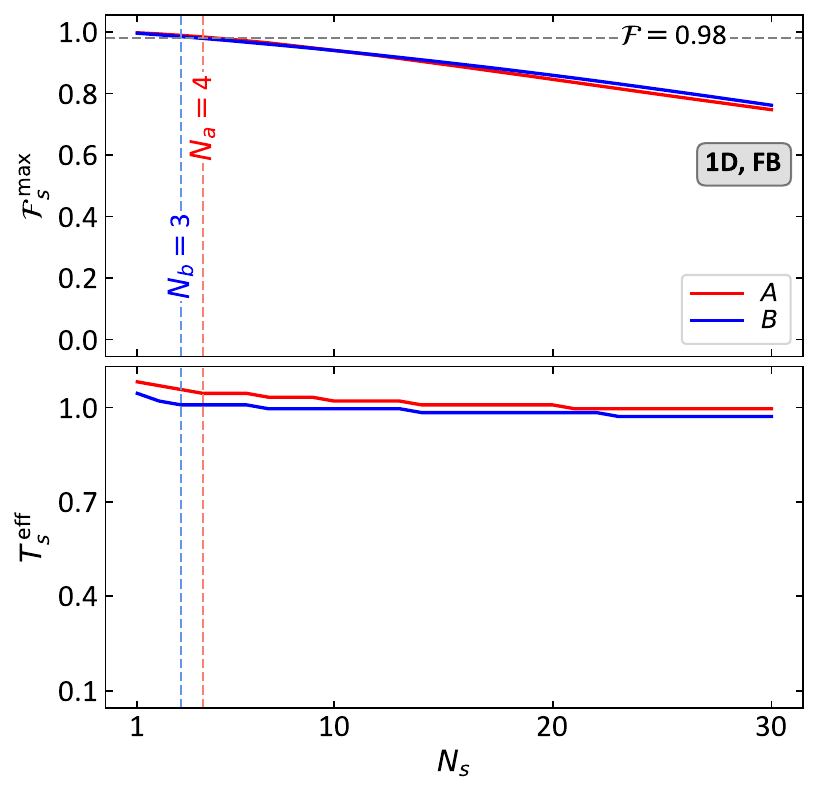}
\caption{
(\textbf{{\Glocal} thermality with subsystem size $N_s$.}) \textbf{Top panel}: the degree of {\glocal} thermality~\eqref{fideldef} of $X$ at an $N_s$-site subsystem $s$ for $N_s = 1, 2, \dots$. All subsystems $s$ are centered within $X$, and all the other parameters are as in Figs.~\ref{fig:global_atherm_1DFB}--\ref{fig:gloc_site_1DFB}. The plot is taken at the instance $\sqrt{\omega_A \omega_B} \, t_0 = 106$. $\cF^{\max}_s$ is high ($\geq 0.98$) up to subsystem sizes $N_a=4$ and $N_b=3$ for these 1D chains of $N_A=N_B=200$. \textbf{Bottom panel}: the effective {\glocal} temperatures $T^{\eff}_s$ \eqref{teffdef} vs $N_s$. These values can be considered trustworthy only up to, respectively, $N_a = 4$ and $N_b = 3$. Interestingly, one finds that they do not change significantly as the subsystem size goes beyond $N_s = 4$.}
\label{fig:gloc_subsize_1DFB}
\end{figure}

\begin{figure}[!t]
\centering
\includegraphics[width = \columnwidth]{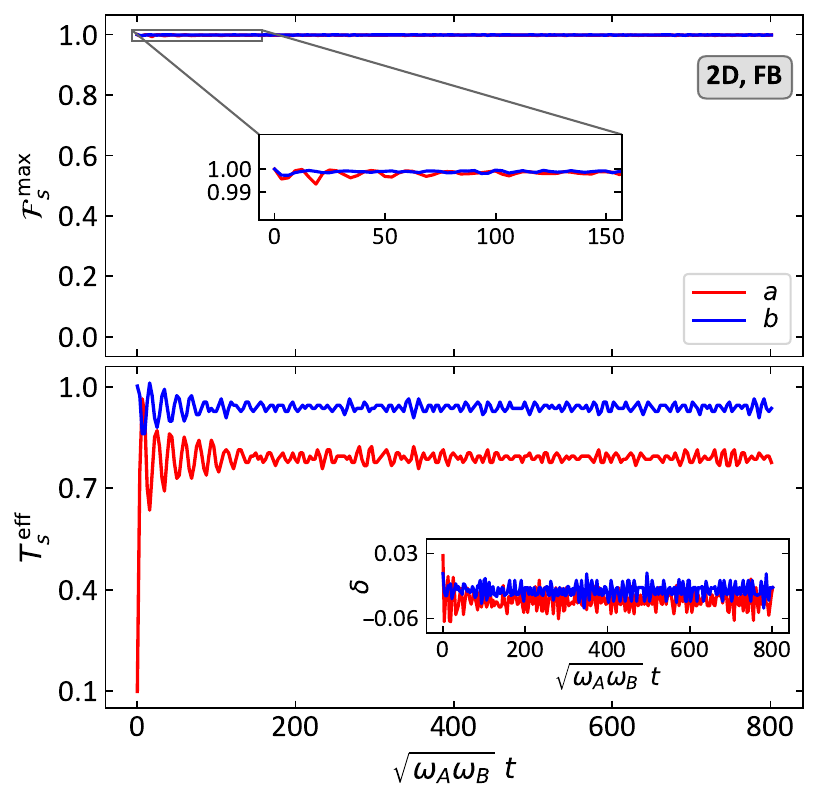}
\caption{
(\textbf{{\Glocal} thermality for [2D, FB].}) 
Analog of Fig.~\ref{fig:gloc_time_1DFB} for a [2D, FB] system (bottom-right corner of Fig.~\ref{fig:couplings}). $A$ and $B$ are $26 \times 26$ 2D lattices, 
with nearest neighbor coupling within them [$\alpha = \infty$ in Eq.~\eqref{decayrate}]), and full-body coupling between them. The subsystems $a$ and $b$ are $2$-site subsystems located centrally in $A$ and $B$, respectively.
We observe very high degrees of {\glocal} thermality in both $A$ and $B$ for all times, as well as faster temperature stabilization than in Fig.~\ref{fig:gloc_time_1DFB}. The inset shows the relative temperature difference $\delta$ between $T_X^\eff$ and $T_X^{\eff, \can}$.
Similarly to the 1D case, the subsystems $a$ and $b$ settle to slightly different {\glocal} temperatures.
Here the Hamiltonian parameters are $\omega_A = 2$ and $g_A/\omega_A^2 = 0.15$ for $A$, $\omega_B = 1.5$ and $g_B/\omega_B^2 = 0.2$ for $B$, and the inter-chain coupling is $\lambda/(\omega_A \omega_B) = 0.23$. The initial temperatures are $T_A = 0.1$ and $T_B = 1$.
}
\label{fig:gloc_time_2DFB}
\end{figure}

Of course, as the size of the subsystem $s$ grows, the degree to which the system $X$ is {\glocal}ly thermal at $s$, $\cF^{\max}_s$, must decrease, reaching very low values as $N_s$ approaches $N_X$ (see Fig.~\ref{fig:global_atherm_1DFB} which plots the fidelity for $N_s = N_X$). The decrease of $\cF^{\max}_s$ with $N_s$ is shown in Fig.~\ref{fig:gloc_subsize_1DFB}, where the center of subsystem $s$ is fixed at the center of $X$. The plot is a snapshot of the system taken at the same instance $\sqrt{\omega_A \omega_B} \, t_0 \approx 106$ as in Fig.~\ref{fig:gloc_site_1DFB}. The presented behavior is representative for all times. We see that $\cF^{\max}_s \geq 0.98$ for $N_a \leq 4$ and $N_b \leq 3$. The effective {\glocal} temperatures $T^\eff_s$ for all values of $N_s$ are approximately equal.
For larger values of time as well as for larger sizes of the global systems, both curves in Fig.~\ref{fig:gloc_subsize_1DFB} become flatter. However, they do not become entirely flat in the $N_X \to \infty$ limit for all times. Thus, although $A$ and $B$ are indeed {\glocal}ly thermal to a good approximation at all times for $N_X \gg 1$, Eq.~\eqref{eq:exact} does not become \textit{exact} in the thermodynamic limit (at least not for all times).

The {\glocal} thermality we observe is not limited to the [1D, FB] case. We find qualitatively identical behavior for the other three topological configurations [1D, EE], [2D, EE], [2D, FB] (see Fig.~\ref{fig:illustr}).
To provide representative evidence, in Fig.~\ref{fig:gloc_time_2DFB} we show the time dependence of the fidelities and effective {\glocal} temperatures for central $2$-site subsystems $a$ and $b$ for the [2D, FB] case. There, $A$ and $B$ are 2D lattices of dimension $26 \times 26$ (i.e., $N=676$) with full-body interaction.
These [2D, FB] plots show qualitative similarity to those for the [1D, FB] case shown in Fig.~\ref{fig:gloc_time_1DFB}, with slightly better convergence compared to the 1D case. Both Figs.~\ref{fig:gloc_time_1DFB} and \ref{fig:gloc_time_2DFB} illustrate the important possibility of the {\glocal} temperatures of $A$ and $B$ not converging to the same value (cf.~Sec.~\ref{sec:TTM}). This behavior can occur both in strong and weak coupling regimes.

For [1D, EE] and [2D, EE] configurations, we found that, while both $A$ and $B$ remain {\glocal}ly thermal, the systems expectedly exhibit gradients of local temperatures. We discuss this in Appendix~\ref{app:EEdyn}.

\smallskip

\textbf{Independence from typicality.---}Finally, the novelty and unexpectedness of our all-time {\glocal} thermality result is emphasized by that it applies to systems and situations well beyond the scope of all known results in canonical typicality and ensemble equivalence. Indeed, the most general result in that direction is the stronger ensemble equivalence proven by Brand\~{a}o and Cramer \cite{Brandao_2015} for lattices with short-range interactions. There it is shown that, if $\tau\big( T_X^{\eff, \can}, H_X \big)$ has exponentially decaying correlations and $\rho_X$ is not too far from $\tau\big( T_X^{\eff, \can}, H_X \big)$, then $\rho_s$ approaches $\tau_s^\MF\big( T_X^{\eff, \can} \big)$ in the thermodynamic limit for most small subsystems $s$. In our language, this means that $\rho_X$ is {\glocal}ly thermal with uniform {\glocal} temperature $T_X^\eff = T_X^{\eff, \can}$. While the conditions under which this result applies are fairly restrictive, especially when dealing with dynamical states, it implies that our result for FB-coupled nearest-neighbor lattices could be expected to some extent. And indeed, in the bottom inset of Fig.~\ref{fig:gloc_time_2DFB}, we see that the difference between $T_X^\eff$ and $T_X^{\eff, \can}$ remains fairly small at all times. However, long-range interacting systems, as well as situations with temperature gradient, are beyond the scope of Ref.~\cite{Brandao_2015} (and all other works on canonical typicality and ensemble equivalence known to us). Sure enough, we find a significant discrepancy between $T_X^\eff$ and $T_X^{\eff, \can}$ in Figs.~\ref{fig:gloc_time_1DFB} and \ref{fig:gloc_time_1DEE}, signalling that the results of Ref.~\cite{Brandao_2015} do not hold. The persistent {\glocal} thermality we observe, on the other hand, applies to all these systems and situations, which strongly suggests that it is a phenomenon fundamentally different from ensemble equivalence. We provide further context and a more detailed discussion in Appendix~\ref{app:brandrammer}.

\section{Equilibration and the generalized Gibbs ensemble}
\label{sec:equil}

Recall that all-time {\glocal} thermality of $A$ and $B$ is guaranteed whenever \textit{all local states} of $AB$ equilibrate at long times. Here we first discuss the details of this requirement and then describe how the equilibrium state relates to the generalized Gibbs ensemble (GGE).

\begin{figure}[!t]
\centering
\includegraphics[width=\columnwidth]{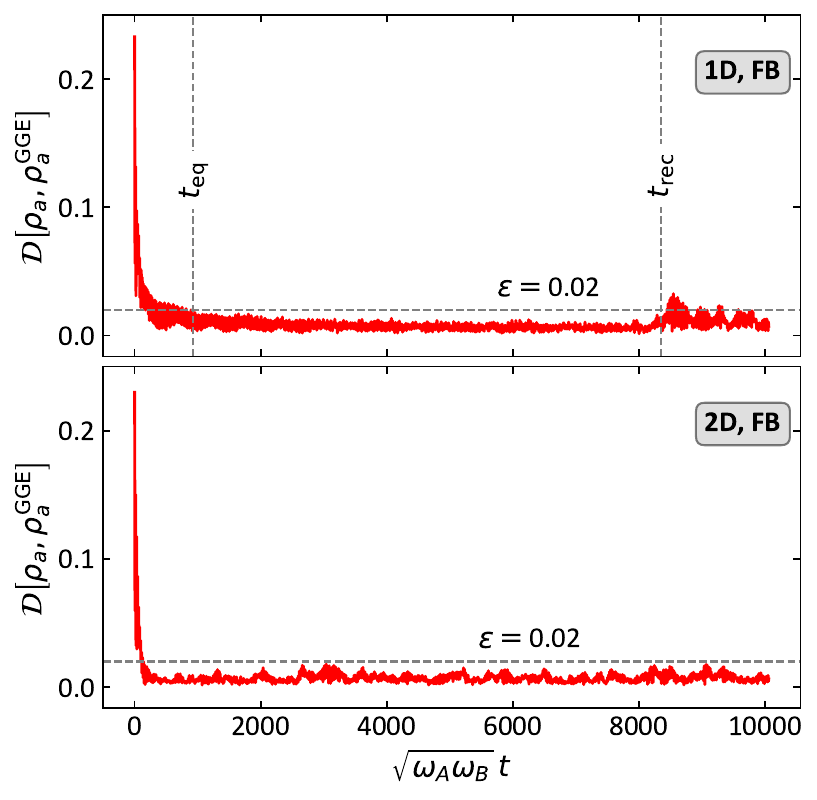}
\caption{
(\textbf{Equilibration and timescales.})
\textbf{Top panel:} Equilibration of a two-site subsystem $a$ in the center of $A$ for [1D, FB] configuration with $\alpha = 1.75$ and all the other parameters as in Fig.~\ref{fig:gloc_time_2DFB}. Here the normal-frequency spectrum is nondegenerate and therefore the equilibrium state $\rho_a^\equi$ equals $\rho_a^\GGE$, with the latter defined in Eq.~\eqref{equil}. Due to the finite size of $AB$, equilibration never occurs exactly---one usually fixes a small $\epsilon > 0$ and considers the system equilibrated once $\cD[\rho_a(t), \rho^\equi] \leq \epsilon$. We choose $\epsilon = 0.02$ for this panel, and find that $s$ is in equilibrium for $t \in [\trel, \trev]$. For times larger than the ``recurrence time'' $\trev$, $AB$ deviates from local equilibrium. This $\trev$ grows with the size of $AB$.
\textbf{Bottom panel:} Equilibration of a two-site subsystem at the center of $A$ for [2D, FB] configuration with $\alpha = 1$ and all the other parameters as in Fig.~\ref{fig:gloc_time_2DFB}. The equilibrium is still described by the GGE, but one that is complemented with the additional conserved charges present in this configuration due to degeneracies in the normal-mode spectrum.
}
\label{fig:equilibration}
\end{figure}

The definition of ``local equilibration'' \cite{Linden_2009, Gogolin_2016} of the total system $AB$ is that the reduced state $\rho_s(t) = \tr_{AB \backslash s} [ \rho_{AB}(t)]$ of each small subsystem $s$ of $AB$ reaches an $\epsilon$-neighbourhood of some $\rhoseq$ within some finite time $\trel(\epsilon)$, and thereafter stays in it. Typically, $\trel(\epsilon)$ will depend only weakly on the size of $AB$, as long as $AB$ is large enough. Note though that, in general, the larger the relative size $N_s/N_{AB}$ of $s$, the larger $\epsilon$ one has to tolerate; see also the related discussion at the end of Sec.~\ref{sec:glocal}.

On the other hand, for a finite system of size $N_{AB}$, there is always an upper time limit, the recurrence time $\trev$, at which the local equilibration behavior is disrupted and information starts flowing back from $AB \backslash s$ into the subsystem $s$. This timescale is typically a monotonically increasing function of $N_{AB}$. Hence, ``local equilibration'' of $AB$ refers to being in equilibrium in the full time-interval $[\trel, \trev]$ \cite{Cramer_2010}. Numerically, we confirm local equilibration by directly calculating $\rho_s(t)$ from $t=0$ to some large $t_{\max}$ and plotting the distance $\cD[\rho_s(t), \rho_s^\GGE]$ against $t$. When this distance goes below some small $\epsilon$ and stays there for a substantial portion of the time, then we conclude that $\rho_s^\equi = \rho_s^\GGE$ (for a more precise characterization, see Appendix~\ref{app:quantifex}). This procedure is done for all small $s$.

Representative results are shown in Fig.~\ref{fig:equilibration} for [1D, FB] and [2D, FB] configurations. We emphasize again that, when such local equilibration of $AB$ occurs at long times, $t > \trel$, then {\glocal} thermality of $A$ and $B$ holds \textit{at all times}, including the transient time interval $[0, \trel]$.

\medskip

Let us now comment on the nature of the equilibrium state itself. For an integrable system (the total system $AB$ in our case), whenever local equilibration takes place, it is generically described by the so-called generalized Gibbs ensemble (GGE) \cite{Kinoshita_2006, Rigol_2007, Rigol_2008, Barthel_2008, Polkovnikov_2011, Gring_2012, Fagotti_2014, Langen_2015, Essler_2016, Gogolin_2016, Vidmar_2016, Tang_2018, Kranzl_2023}.
For systems with quadratic Hamiltonians (bosonic and fermionic alike), the GGE is determined only by quadratic conserved charges \cite{Barthel_2008, Murthy_2019, Gluza_2019}. Whenever all the normal frequencies of the system are different from each other, the Hamiltonians of the normal modes, which are conserved, constitute a basis in the algebra of conserved charges.
Thus, when all $N_{AB}$ normal frequencies $\Omega_k$ of the interacting harmonic lattice $AB$ are distinct, the GGE takes the form
\bea \label{def:GGE}
	\rho^\GGE_{AB} := \frac{e^{- \sum_\kappa \beta_\kappa h_\kappa}}{\tr[e^{- \sum_\kappa \beta_\kappa h_\kappa}]} ,
\eea
where $h_\kappa := \frac{\Omega_\kappa}{2} (Q_\kappa^2 + P_\kappa^2)$ are the normal-mode Hamiltonians ($Q_\kappa$ and $P_\kappa$ being the normal-mode coordinates) of $AB$. By definition, the total post-quench Hamiltonian can be written as $ H_\tot = \sum_{\kappa = 1}^{N_{AB}} h_\kappa$ (see Appendix~\ref{app:Gaussian}).

The state \eqref{def:GGE} describes the equilibrium in the sense that \cite{Barthel_2008, Cramer_2010}
\bea \label{equil}
	\trel \leq t \leq \trev: \;\;\;
	\rho_s (t) \approx \tr_{AB \backslash s} \big[\rho_{AB}^\GGE\big] := \rho_s^\GGE, ~~
\eea
where the approximate equality sign indicates that there will generally be a finite-size correction $\epsilon$. In Eq.~\eqref{def:GGE}, $1/\beta_\kappa$ are the ``generalized temperatures'' that stem from the fact that the $h_\kappa$'s are conserved in the dynamics. They are determined through the initial expectation values 
\bea \label{betakappa}
    \tr \big[h_\kappa \, \rho_{AB}^\GGE\big] = \tr [h_\kappa \, \rho_{AB}(0)], \;\;\;
    \kappa = 1, \cdots, N_{AB} ~~~
\eea
(see Eq.~\eqref{ugly} for an explicit formula). The top panel of Fig.~\ref{fig:equilibration} shows local convergence to this GGE for [1D, FB] configuration.

When the spectrum of normal frequencies has degeneracies, the $h_\kappa$'s no longer span the complete algebra of conserved charges \cite{Fagotti_2014, Murthy_2019, Gluza_2019}. More specifically, each pair $\Omega_k = \Omega_j$ ($k \neq j$) adds the conserved charge $I_{kj} = \Omega_k (Q_k Q_j + P_k P_j)$. Together with the $h_\kappa$'s, these now span the complete algebra of conserved charges. Therefore, in order to correctly describe the system's local equilibrium, the GGE needs to be complemented accordingly: $\rho^\GGE_{AB} \propto e^{-\sum \beta_\kappa h_\kappa - \sum \beta_{kj} I_{kj}}$ \cite{Fagotti_2014, Murthy_2019, Gluza_2019}. Similarly to Eq.~\eqref{betakappa}, the $\beta_{kj}$'s are determined from $\tr[I_{kj} \rho^\GGE_{AB}] = \tr[I_{kj} \rho_{AB}(0)]$. Due to the presence of degeneracies, the decomposition of $H_{AB}$ into normal modes is not unique. Conveniently, one can always choose a set of normal modes $(\widetilde{Q}_\kappa, \widetilde{P}_\kappa)$ such that all $\tr[\widetilde{I}_{kj} \rho_{AB}(0)]=0$, which in turn leads to $\widetilde{\beta}_{kj} = 0$ \cite{Murthy_2019}. With such a choice of normal modes, the GGE again takes the form \eqref{def:GGE}, now depending on the $\widetilde{I}_{kj}$'s only indirectly, through the conditions $\tr[\widetilde{I}_{kj} \rho_{AB}(0)]=0$. We follow this procedure in our numerics whenever the system has a degenerate normal frequency spectrum.

In our numerical experiments, only the [2D, FB] configuration yielded degenerate normal frequency spectra. In all other configurations, the spectrum was always nondegenerate. This might be related to the fact that the [2D, FB] is the only configuration for which $AB$ is effectively three-dimensional (cf.~Fig.~\ref{fig:couplings}).

That the local equilibrium of harmonic systems is described by the GGE has been established in the literature for the following two scenarios: (i) Normal frequency spectrum must be nondegenerate, but the range of interactions can be arbitrary \cite{Barthel_2008}; (ii) normal frequency spectrum can be degenerate, but the interactions must be of finite range or decaying exponentially \cite{Fagotti_2014, Murthy_2019, Gluza_2019}. Sure enough, our numerics confirms that the GGE describes the equilibrium for [1D, EE], [1D, FB], [2D, EE] $\forall \alpha$ (the $h_\kappa$'s are sufficient) and for [2D, FB] with $\alpha = \infty$ (the $I_{kj}$'s have to be accounted for).

However, the configuration [2D, FB] with $\alpha < 2$, where the normal-frequency spectrum is degenerate and the interactions are of long range, is not covered by any of the known results about harmonic systems. For this case, we establish that the equilibrium is still described by the GGE that accounts for the charges $I_{kj}$. The bottom panel of Fig.~\ref{fig:equilibration} illustrates such a situation on the example of a [2D, FB] lattice with long-range interactions ($\alpha = 1$).


\section{Two-temperature model (TTM) and {\glocal}ity}
\label{sec:TTM}

Let us now discuss the implications of {\glocal} thermality for the TTM in the strong-coupling regime. The two-temperature model is widely used in solid-state physics \cite{Anisimov_1967, Anisimov_1974, Sanders_1977, Allen_1987, Jiang_2005, Carpene_2006, Lin_2008, Liao_2014} to describe a setting similar to ours. Namely, it concerns the joint thermalization of two macroscopic systems that start at different temperatures. Usually one of the systems, say $A$, is a free electron gas while the other, $B$, is a crystal lattice. However, the TTM is not specific to those systems and can be formulated generally, based on two assumptions.
 
First, the TTM posits that each system $A$ and $B$ can be described by a thermal state at all times. In our notation that would mean that the reduced states must be global Gibbs states throughout, $\rho_A (t) = \tau(T_A(t), H_A)$ and $\rho_B (t) = \tau(T_B(t), H_B)$, where the temperatures $T_A(t)$ and $T_B(t)$ vary in time. 
The second assumption made by the TTM (and many of its generalizations) concerns the energy exchange between $A$ and $B$. It assumes that the energy exchange is governed by a rate equation, with rates given by a Fourier-like law \cite{Anisimov_1967, Anisimov_1974, Sanders_1977, Allen_1987, Jiang_2005, Lin_2008, Wang_2012, Liao_2014}. In Appendix~\ref{app:monotonic}, we write this rate equation explicitly and show that its validity is equivalent to the assumption that the temperatures of the systems, $T_A(t)$ and $T_B(t)$, are differentiable functions of time that converge monotonically.
Thus, the second assumption can be neatly summarized as ``$T_A(t)$ and $T_B(t)$ monotonically approach the same value $T^\equi$.''
The TTM's standard regime of validity is when $A$ and $B$ interact relatively weakly, whereas at strong couplings it is known to break down \cite{Waldecker_2016}. 

Two key applications of the TTM are noteworthy here. 
First, it allows one to determine the equilibrium temperature $T^\equi$ \cite{Anisimov_1967, Anisimov_1974, Sanders_1977, Reppert_2016, Pudell_2018, Herzog_2022} which is fixed by energy conservation, i.e.,
\beaa \label{blimp}
	\langle H_A \rangle_{T_A} + \langle H_B \rangle_{T_B} = \langle H_A\rangle_{T^\equi} + \langle H_B \rangle_{T^\equi} ,
\eeaa
where $\langle H_X \rangle_{T} := \tr[H_X \, \tau(T, H_X)]$ and $T_X$ is the initial temperature of $X$. The lack of accounting for the energy stored in the interaction $H_\I$ is a manifestation of the weak coupling assumption.
Second, the rate equation allows inferring the temporal changes of the temperatures and energies of the interacting systems $A$ and $B$ \cite{Anisimov_1974}. This has been used to understand the ablation of metals following ultra-short pulses \cite{Byskov-Nielsen_2011} and to characterize the ultrafast heat transport between electrons and phonons in multi-layers \cite{Pudell_2018}. Extensions to a three-temperature model which includes their interaction with spins have proven useful in the study of ultrafast demagnetization processes \cite{Beaurepaire_1996, Zhang_2002, Kazantseva_2008}.

Based on our findings for harmonic lattices, we can now comment on the validity of the TTM beyond the weak coupling limit it was originally intended for. 
The electron--phonon setting of the original TTM corresponds to the FB coupling scenario in our setup.
In Fig.~\ref{fig:global_atherm_1DFB}, we see that both $A$ and $B$ move very quickly away from Gibbs states, even at fairly weak couplings. Hence, they have no well-defined global temperatures.
This breakdown of all-time global Gibbsianity beyond the extremely weak coupling regime is not unexpected \cite{Kazantseva_2008, Waldecker_2016, Maldonado_2017}.

\begin{figure}[!t]
\centering
\includegraphics[width = \columnwidth]{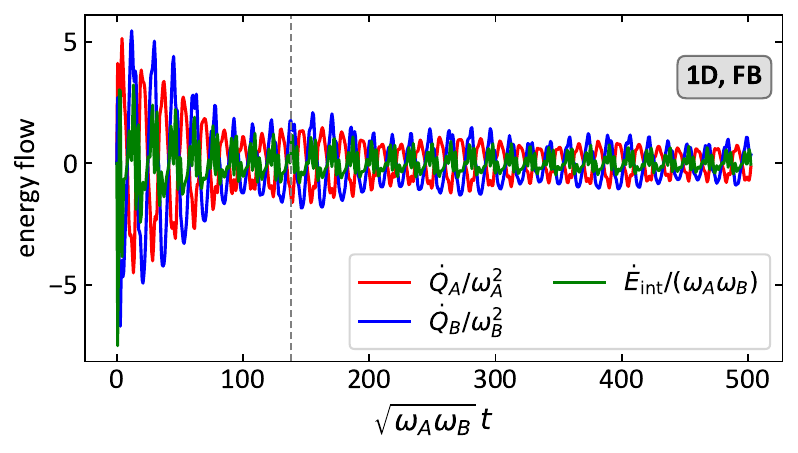}
\caption{
(\textbf{Energy flows.})
Heat flows to each global system, $\dot{Q}_X := d\tr[H_X \rho_X]/dt$, and interaction energy flow, $\dot{E}_\I$, as functions of time. 
All parameters are as in the top panel of Fig.~\ref{fig:equilibration}.
Note that the direction of heat flow from $A \to B$ and from $B \to A$ is
oscillatory. Moreover, due to the eventual equilibration of the whole system, all energy flows slowly converge to near-zero values as time goes on beyond what is shown in the plot.
This plot represents generic behavior for all four geometries in Fig~\ref{fig:couplings}.
The vertical dashed line is chosen at $\sqrt{\omega_A \omega_B} \, t_{\mathrm{bf}} \approx 137.87$ where heat backflow occurs: $A$ has a lower temperature than $B$, but loses heat ($\dot{Q}_A < 0$) while $B$ receives heat ($\dot{Q}_B > 0$).
%
}
\label{fig:heatflow}
\end{figure}

What is perhaps surprising is that we here find that \textit{it is possible} to associate {\glocal} temperatures \eqref{teffdef} with $A$ and $B$ at \textit{all times}; see Figs.~\ref{fig:gloc_time_1DFB} and \ref{fig:gloc_time_2DFB}. In this sense, the first assumption of the TTM can be rescued at strong coupling. Moreover, our finding that all-time {\glocal} thermality also holds in the presence of temperature gradients (see Appendix~\ref{app:EEdyn}) opens the possibility of upgrading even the more general diffusive TTM \cite{Anisimov_1974, Lin_2008, Pudell_2020, Herzog_2022} to the strong coupling regime. The latter posits ``local thermal equilibrium'' within each lattice, i.e., that each small, localized subsystem of the lattice is in a Gibbs state with respect to its own bare Hamiltonian \cite{Cahill_2003, Herzog_2022}. Of course, when there is strong coupling within the lattice, the local thermality hypothesis breaks down, whereas the {\glocal} thermality is maintained.

Regarding the second assumption of the TTM, we find that its main proposition does not hold anymore for harmonic lattices even for the extended notion of {\glocal} temperatures. This is evidenced by Figs.~\ref{fig:gloc_time_1DFB} and \ref{fig:gloc_time_2DFB} which clearly show that $T^\eff_A(t) - T^\eff_B(t)$ is not monotonic in $t$. Therefore, beyond the weak coupling limit,
no TTM-type rate equation exists that would describe the time evolution of $T^\eff_A$ and $T^\eff_B$. This is so despite the fact that heat capacities are well-defined for both $A$ and $B$ because they are {\glocal}ly thermal (see the discussion in Appendix~\ref{app:local_meas}). 

Nevertheless, we find that the predictive power of Eq.~\eqref{blimp} is partially retained for harmonic lattices.
%
Indeed, although $T^\eff_A$ and $T^\eff_B$ may in general converge to two different values (see Figs.~\ref{fig:gloc_time_1DFB} and \ref{fig:gloc_time_2DFB}), Eq.~\eqref{blimp} remains fairly accurate with $T^\equi$ substituted by $T^{\eff, \equi}_A$ and $T^{\eff, \equi}_B$. See a detailed discussion on this in Appendix~\ref{app:energycondition}.

Lastly, we note that, together with the non-monotonic convergence of temperatures, the alternating direction of the heat flow shown in Fig.~\ref{fig:heatflow}, witnesses (but does not necessitate \cite{Schmidt_2016}) the non-Markovian nature of the dynamics each system $X$ is undergoing under the influence of the other. Moreover, in contrast to the predictions of the TTM, energy may sometimes flow from cold to hot, a phenomenon sometimes referred to as ``backflow of heat.''

\section{Discussion}
\label{sec:discussion}

To summarize, going beyond the too restrictive demand of global thermality, we have introduced the notions of {\glocal} thermality and the associated {\glocal} temperatures.
These characterise local subsystems while also making reference to global Gibbs states of the many-body system.
We have evidenced the power of these concepts on the example of a pair of harmonic lattices with varying spatial dimensions and topologies of couplings.
We found compelling numerical evidence that persistent {\glocal} thermality of $A$ and $B$ at transient times is a necessary condition for $AB$ to thermalize at long times. 
This is true even though $A$ and $B$ themselves venture far from being globally thermal, and applies to lattices $A$ and $B$ with both short and long range interactions within them, as well as arbitrary coupling strengths between them.
This finding adds a new ``expected behavior'' to the short list of known results for the transient regime in the dynamics of interacting quantum many-body systems.
Furthermore, for the equilibrium state itself, we found that it is described by the GGE for all configurations and interaction ranges. This includes the peculiar case of 2D lattices with full-body coupling (rendering $AB$ three-dimensional), for which the normal-frequency spectrum is degenerate. Such systems have an extended algebra of conserved charges, and the GGE has to be constructed taking into account all those charges.

These results open several new directions. As a first step, many-body systems other than harmonic lattices may be tested numerically for the presence of {\glocal} thermality. Further ahead, analytical arguments might be constructed that can prove the presence of transient {\glocal} thermality given long-term equilibration for either harmonic lattices or more general many-body systems.
Finally, experiments with atoms in optical lattices or trapped ions may in the future test the link between transient {\glocal} thermality and long-time equilibration \cite{Kinoshita_2006, Bendkowsky_2009, Gring_2012, Britton_2012, Langen_2015, Gross_2017, Bernien_2017, Neyenhuis_2017, Tang_2018, Kranzl_2023}.

In general, {\glocal} thermality naturally fits into the framework of quantum thermometry \cite{Pasquale_2018, Mehboudi_2019} and strong coupling thermodynamics \cite{Seifert_2016, Jarzynski_2017, Miller_2018Rev, Trushechkin_2022}.
Performing spatially local thermometry \cite{Pasquale_2016, Campbell_2017, Hovhannisyan_2018, Pasquale_2018, Mehboudi_2019} measures a system's {\glocal} temperature irrespective of whether the system state is globally thermal or {\glocal}ly thermal (see Appendix~\ref{app:local_meas}). Such measured records give an operational meaning to {\glocal} temperatures.
Moreover, when dealing with many-body systems with local Hamiltonians, all energetic quantities are already determined by local states. Thus, those strong-coupling thermodynamics results which are derived under the assumption of global thermality \cite{Miller_2018NC, Perarnau-Llobet_2018, Hovhannisyan_2020, Trushechkin_2021, Henkel_2021, Anto-Sztrikacs_2022}, will naturally extend to {\glocal}ly thermal systems. Furthermore, all-time {\glocal} thermality may enable hydrodynamic treatment of nonequilibrium transport at strong coupling not only in the steady state \cite{Castro-Alvaredo_2016, Bulchandani_2017} but also in the transient. {\Glocal} thermality may also be useful in the study of local transfer in systems with non-commuting conserved charges \cite{Manzano_2022, Majidy_2023}. Lastly, maintained {\glocal} thermality might lead to a type of Markovianity and local detailed balance for some observables \cite{Strasberg_2023} under certain conditions.

Part of the motivation for this work was to provide a microscopic justification of the two-temperature model (TTM) often used to interpret transient heat dynamics in condensed-matter systems. 
The TTM assumes that both systems remain globally thermal during the interaction, an assumption that generally fails when the coupling is not weak. For our model system, we saw that some of the features of the TTM can be carried over into the strong coupling regime, by updating the restrictive global thermality assumption to {\glocal} thermality.
However, at strong couplings, we saw that the {\glocal} temperatures of $A$ and $B$ relax in an oscillatory fashion, and that their difference may remain nonzero. This behavior is clearly incompatible with a rate equation ansatz for heat exchange typically applied within the TTM. Nonetheless, the maintenance of {\glocal} thermality and the ability to write a simple (approximate) energy condition for the equilibrium ({\glocal}) temperatures akin to Eq.~\eqref{blimp} provides a substantial generalization of the TTM to the strong coupling regime.
The phenomenology we found for harmonic lattices may admittedly not be fully transposed to ``hot electrons'' exchanging heat with a ``cold crystal lattice.'' The approach we propose is however flexible enough to capture both kinds of subsystems one typically encounters in condensed-matter physics: either localized in a limited spatial domain or defined by a certain set of (coarse-grained) physical observables.

\medskip

\noindent\textbf{Acknowledgments.}
We thank Philipp Strasberg for interesting discussions.
K.H., S.N., and J.A. are grateful for support from the University of Potsdam.
J.A. gratefully acknowledges funding from the Deutsche Forschungsgemeinschaft (DFG, German Research Foundation) under Grants No. 384846402 and No. 513075417 and from the Engineering and Physical Sciences Research Council (EPSRC) (Grant No. EP/R045577/1) and thanks the Royal Society for support.
Open access publication is funded by the Deutsche Forschungsgemeinschaft (DFG, German Research Foundation), Project No. 491466077.

\bibliography{references}

\appendix
\setcounter{corollary}{0}
\renewcommand{\thecorollary}{\thesection.\arabic{corollary}}

\section{FIDELITY AND BURES DISTANCE}
\label{app:dist}

In definition \eqref{teffdef}, one can in principle use any metric to define the effective temperature. With any choice of metric, the resulting effective temperature will coincide with the true {\glocal} temperature whenever the system is {\glocal}ly thermal exactly.

Since our model is Gaussian (see Appendix~\ref{app:Gaussian}), it is convenient to work with the Bures metric. It is defined as \cite{mikeike}
\bea\label{eq:distance}
	\cD(\rho_1, \rho_2) := \big[ 2( 1 - \sqrt{\cF(\rho_1, \rho_2)}) \big]^{1/2},
\eea
where
\bea
	\mathcal{F}(\rho_1, \rho_2) := \Big( \tr \sqrt{\rho_1^{1/2} \rho^{\phantom{1/2}}_2 \!\!\! \rho_1^{1/2}} \Big)^2
\eea
is the quantum fidelity \cite{mikeike}. The reason for this preference in our case is that the fidelity can be explicitly calculated for Gaussian multimode states through their covariance matrices \cite{Scutaru_1998, Marian_2012, Banchi_2015}; see Eqs.~\eqref{GFidel_1}--\eqref{GFidel_3}.

\section{G-LOCAL THERMALITY AND LOCAL OBSERVABLES}
\label{app:local_meas}

Let us show that, by using only local observables, one cannot differentiate between standard (globally) thermal and uniformly {\glocal}ly thermal states of many-body systems.

An observable $O$ living in the Hilbert space of a lattice system $X$ is called local if it can be written as
\bea
	O = \sum_{s \subset X} O_s,
\eea
with each $O_s$ acting nontrivially only on some spatially localized subsystem $s$ containing at most $k$ sites, for some fixed $k$. This means that one can write $O_s = \hat{O}_s \otimes \id_{X\backslash s}$, where $\hat{O}_s$ is some operator living in the Hilbert space of $s$.

Now, if the state of $X$, $\rho_X$, is {\glocal}ly thermal at each $s$, with uniform {\glocal} temperature $T$, then
\beaa \label{locop}
	\langle O \rangle :=& \tr_X [O \rho_X]
\\
	=& \sum_s \tr_X \left[\left(\hat{O}_s \otimes \id_{X \backslash s}\right) \, \rho_X\right]
\\
	=& \sum_s \tr_s \left[\hat{O}_s \, \rho_s\right]
\\
	\stackrel{(*)}{=}& \sum_s \tr_s \Big[\hat{O}_s \tr_{X \backslash s} \big[\tau(T, H_X)\big] \Big]
\\
	=& \sum_s \tr_X \left[\left(\hat{O}_s \otimes \id_{X \backslash s}\right) \, \tau(T, H_X)\right]
\\
	=& \tr_X \left[O \, \tau(T, H_X)\right],
\eeaa
where in step $(*)$ the {\glocal}ity condition \eqref{def:g-local} was used.

This in particular means that, if $H_X$ is local, then the effective canonical temperature $T_X^{\eff, \can}$ of $X$, defined in Eq.~\eqref{Teff_can}, coincides with $T$. Indeed, in view of Eqs.~\eqref{locop}, one has $\tr[\rho_X H_X] = \tr[H_X \tau(T, H_X)]$. Moreover, if we define the ``{\glocal}'' heat capacity of $X$ as $d \tr[H_X \rho_X] / dT$, then it will be equal to the heat capacity of $X$ if it were in a global thermal state at temperature $T$.

Note that, in some cases, it might happen that $\rho_X$ is {\glocal}ly thermal also at small, but spatially \textit{delocalized} subsystems containing at most $k$ sites. Then, the equality $T^{\eff, \can}_X = T^{\eff}_X$ will hold even if $H_X$ is a long-range, but at most $k$-body, interacting Hamiltonian.

Finally, we note that the effective canonical temperature $T_X^{\eff, \can}$ has found some use in nonequilibrium thermodynamics \cite{Swendsen_2015, Seifert_2020, Strasberg_2021}, and the equality $T^{\eff, \can}_X = T^{\eff}_X$ for {\glocal}ly thermal systems provides $T_X^{\eff, \can}$ with an additional thermodynamic meaning.

\section{A SUMMARY OF HARMONIC SYSTEMS}
\label{app:Gaussian}

As described in Sec.~\ref{sec:setup}, our model system is a harmonic lattice. Namely, it is a collection of linearly coupled oscillators, so all the Hamiltonians are quadratic. The tools for simulating and calculating many physical and information-theoretical quantities for such systems are well-known and thoroughly described in, e.g., Refs.~\cite{Adesso_2007, Anders_2008, Weedbrook_2012}. Here we will give a brief account of the main notions and formulas necessary for our purposes.

The position and momentum coordinates of a system of $N$ oscillators are conveniently collected into a column vector in the phase space
\bea
	\mathbf{x} = \Big( \begin{array}{c}
	\mathbf{q} \\ \mathbf{p}
	\end{array} \Big),
\eea
with a total of $2N$ components
\bea \nonumber
	\mathbf{q} = (q_{1}, \dots, q_{N})^\T, 
	\qquad
	\mathbf{p} = (p_{1}, \dots, p_{N})^\T.
\eea
We call this phase-space basis the ``q--p'' basis. In this basis, and in units where $\hbar = 1$, the canonical commutation relations are written as
\bea \label{CCR}
	[{x}_{k}, {x}_{j}] = i \Upsilon_{k,j},
\eea
where the antisymmetric matrix $\Upsilon$ has the symplectic form:
\bea \label{def:symplectic-form}
	\Upsilon = \bigg( \! \begin{array}{cc} 0 & \id \\ - \id & 0 \end{array} \! \bigg),
\eea
with $\id$ being the $N \times N$ identity matrix and $0$ the $N \times N$ zero matrix. (A relation similar to \eqref{CCR} applies in classical mechanics, but with the Poisson bracket instead of the commutator.)

In terms of the phase-space coordinates $\mathbf{x}$, the Hamiltonian is a quadratic form
\bea \label{matrixF}
	H = \frac{1}{2} \mathbf{x}^\T F \mathbf{x},
\eea
where we call the symmetric matrix $F$ the ``Hamiltonian matrix.'' When there is no momentum--momentum coupling in the system, $F$ will take the form
\bea \label{fchi}
	F = V \oplus \id,
\eea
where $V$ corresponds to the ``potential energy'' part of the Hamiltonian, and the identity $\id$ specifies the kinetic energy (after scaling out the oscillator masses).

Two harmonic systems $A$ and $B$ can be combined by forming the direct sum of their phase spaces, with the joint Hamiltonian matrix
\bea
	F_{AB} = V_{AB} \oplus \id_{AB},
\eea
where
\bea
	V_{AB} = V_A \oplus V_B + V_\I.
\eea
Here the interaction potential $V_\I$ represents $H_\I$ that couples $A$ and $B$ [see, e.g., Eq.~\eqref{hint}].

Due to a theorem by Williamson \cite{Williamson_1936}, any Hamiltonian matrix can be symplectically diagonalized. Namely, there exists a symplectic transformation matrix $S$ such that
\bea \label{willi}
	S^\T F S = \Omega \oplus \Omega,
\eea
where the diagonal matrix $\Omega = \diag(\Omega_1, \dots, \Omega_N)$ collects the normal mode frequencies of the system. Recall that symplectic is any matrix that leaves the canonical commutation relation~(\ref{CCR}) invariant: $S^\T \Upsilon S = \Upsilon$.

The same matrix $S$ switches from the q-p basis to the normal mode basis:
\bea \label{norm-orig}
	\mathbf{x} = S \mathbf{X},
\eea
where $\mathbf{X} = (Q_1, \dots, Q_N, P_1, \dots, P_N)^\T$ collects the positions and momenta of the normal modes. In the normal-mode basis, the Hamiltonian is a sum of non-interacting oscillators:
\bea
	H = \sum_{j} \Omega_j \bigg[ \frac{Q_j^2}{2} + \frac{P_j^2}{2} \bigg].
\eea

The initial state $\rho(0)$ \eqref{rhonin} is a tensor product of Gibbs states of quadratic Hamiltonians; therefore, it is Gaussian. Hence, there is no net displacement of the phase-space coordinates, $\langle \mathbf{x} \rangle_t := \tr[\rho(t) \, \mathbf{x} ] = 0$, and, since the Hamiltonian is quadratic at all times, the state $\rho(t)$ also remains Gaussian at all times \cite{Weedbrook_2012}.

Gaussian states are uniquely determined by the covariance matrix \cite{Weedbrook_2012}
\bea \label{covmat}
	\sigma_{jk} = \frac{1}{2} \tr\left[\rho \, (x_j x_k + x_k x_j)\right],
\eea
where the curly brackets denote the anticommutator. Conveniently, the covariance matrix of any subsystem (in our case, it can be $A$ or $B$ or some small subsystem of sites) is simply the corresponding sub-block of $\sigma$, which determines its reduced state.

Whenever the system is in a Gibbs state (i.e., $\rho = \tau(T, H)$ \eqref{eq:globalthermality}), the covariance matrix in the normal-mode basis,
\bea
	\Sigma_{jk} = \frac{1}{2} \tr\left[\tau(T, H) \, ( X_j X_k + X_k X_j ) \right],
\eea
is diagonal and given by \cite{Weedbrook_2012}
\bea
	\Sigma = R \oplus R,
\eea
where $R = \diag(R_1, ..., R_N)$, with
\bea
	R_j = \frac{1}{2} \coth\frac{\Omega_j}{2T} .
\eea
Due to Eq.~\eqref{norm-orig}, the covariance matrix in the ``original'' q--p basis $x$ is then
\bea
	\sigma = S \Sigma S^\T.
\eea
Noting that the covariance matrix for a tensor product $\rho_A \otimes \rho_B$ is a direct sum, $\sigma_A \oplus \sigma_B$, we can thus construct the covariance matrix of the initial state \eqref{rhonin}. 

As for calculating the dynamics of the system, it can be derived immediately from the Heisenberg equations of motion for $\mathbf{x}$ that the evolution of the covariance matrix under a quadratic Hamiltonian is a symplectic transformation \cite{Arnold, Weedbrook_2012}:
\bea \label{gen_evol}
	\sigma(t) = \cE(t) \sigma(0) \cE(t)^\T,
\eea
where $\cE(t)$ is a symplectic matrix. Moreover, $\cE(t)$ is explicitly expressed through the matrix $F$ \eqref{matrixF} \cite{Heffner_1965, Brown_2013}. When the Hamiltonian is time-independent, $F$ is constant and
\bea
	\cE(t) = e^{\Upsilon F t}.
\eea

\medskip

Furthermore, to find effective {\glocal} temperatures through Eq.~\eqref{teffdef}, we need to calculate the fidelity \eqref{fideldef}. For two $N$-mode Gaussian states $\rho_1$ and $\rho_2$ with respective covariance matrices $\sigma_1$ and $\sigma_2$ and identical average coordinates (which are zero in our case), the fidelity is given by \cite{Banchi_2015}
\bea \label{GFidel_1}
	\mathcal{F}(\rho_1, \rho_2) = \sqrt[4]{\frac{M}{\det(\sigma_1 + \sigma_2)}},
\eea
where
\bea \label{GFidel_2}
	M = \det \bigg[2 \bigg(\sqrt{\id + \frac{1}{4}(C \Upsilon)^{-2}} + \id \bigg) C \bigg] ,
\eea
with $\id$ and $\Upsilon$ being the $(2N) \times (2N)$ identity matrix and the symplectic form [Eq.~\eqref{def:symplectic-form}], respectively. The matrix $C$ is defined as
\bea \label{GFidel_3}
	C = - \Upsilon (\sigma_1 + \sigma_2)^{-1} \Big( \frac{\Upsilon}{4} + \sigma_2 \Upsilon \sigma_1 \Big).
\eea

\begin{figure}[!t]
\centering
\includegraphics[width = \columnwidth]{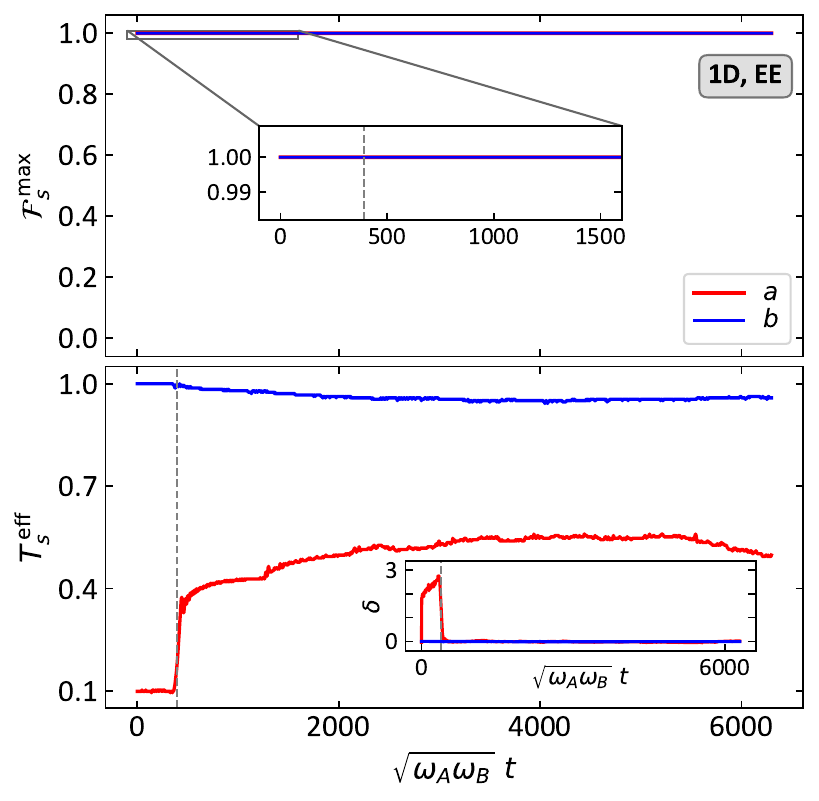}
\caption{(\textbf{{\Glocal} thermality of $A$ at $a$ and $B$ at $b$.}) Similarly to Fig.~\ref{fig:gloc_time_1DFB}, the \textbf{top panel} shows the fidelity $\cF^{\max}_{s}$~\eqref{fideldef} of $A$ at $s=a$ (red) and of $B$ at $s=b$ (blue) as a function of time $t$. For each $X$, the single-site subsystem $s$ is located at the center of the lattice $X$. $A$ and $B$ are connected at $\nu = 1$. The inset zooms into the fidelity at early times. The fidelities remain very close to $1$ (smallest $\cF^{\max}_{s}$ being $0.999979$), indicating that both $A$ and $B$ are {\glocal}ly thermal at $a$ and $b$, respectively, with extremely good precision \textit{at all times} during the evolution.
The \textbf{bottom panel} shows the corresponding effective temperatures $T_{s}^\eff$~\eqref{teffdef}. The inset is for $\delta$---the relative discrepancy between the effective {\glocal} and canonical temperatures. It is large in the transient regime, signalling a significant temperature gradient in the system (cf. Fig.~\ref{fig:gloc_site_1DEE}).
This plot is for $A$ and $B$ both having only nearest-neighbor interactions within them ($\alpha = \infty$). All other parameters are the same as in Fig.~\ref{fig:gloc_time_1DFB}. The vertical dashed line indicates the instance at which Fig.~\ref{fig:gloc_site_1DEE} is plotted.}
\label{fig:gloc_time_1DEE}
\end{figure}

\begin{figure}[!t]
\centering
\includegraphics[width = \columnwidth]{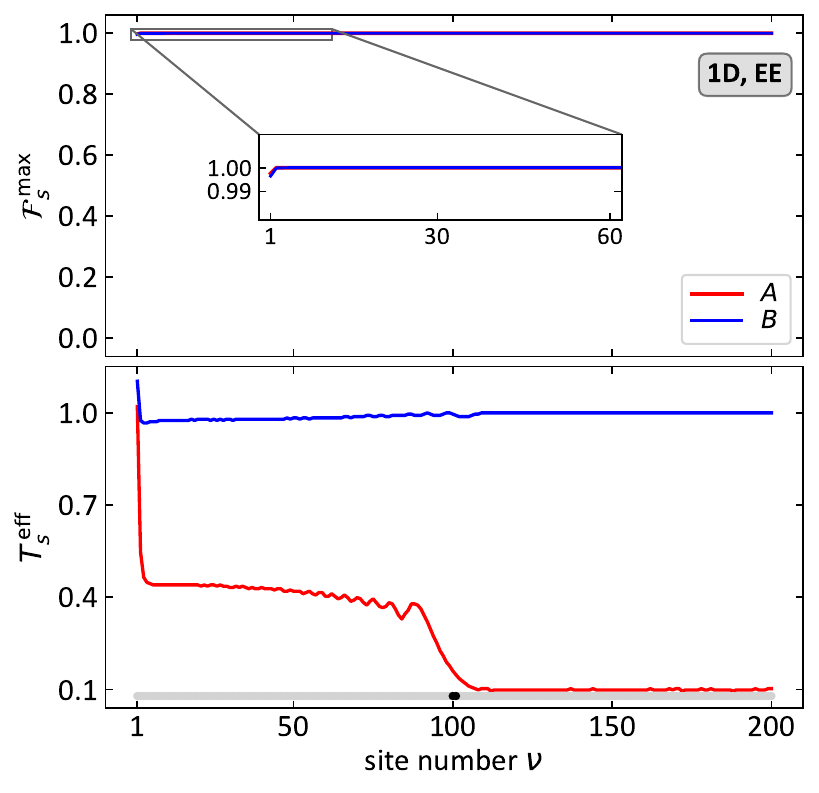}
\caption{
(\textbf{{\Glocal} thermality with site number $\nu$.}) Similarly to Fig.~\ref{fig:gloc_site_1DFB}, the \textbf{top panel} plots the fidelity $\cF^{\max}_s$~\eqref{fideldef} of $A$ at single-site subsystem $s=a$ (red) and of $B$ at $s=b$ (blue), respectively, and the \textbf{bottom panel} shows the corresponding effective temperatures $T_s^\eff$~\eqref{teffdef} as a function of the site number $\nu$. Here $\nu$ labels \textit{all} single-site subsystems $s$ in $X$, with $\nu = 1, ..., N$. 
One sees that the fidelities $\cF^{\max}_{s}$ are very close to 1 for all $\nu$ (the smallest $\cF^{\max}_{s}$ being $0.996439$). However, in contrast to the [1D, FB] case shown in Fig.~\ref{fig:gloc_site_1DFB}, the {\glocal} temperatures $T_{s}^\eff$ \textit{change} with $s$---there is a ({\glocal}) temperature gradient in the system.
The ``slider'' (black dot) at the bottom indicates the position of the subsystem $s$ in $X$ at which Fig.~\ref{fig:gloc_time_1DEE} is plotted.
Here, $\sqrt{\omega_A \omega_B} \, t_0 \approx 389.71$, and all other parameters are as in Fig.~\ref{fig:gloc_time_1DEE}.
}
\label{fig:gloc_site_1DEE}
\end{figure}

\medskip

Lastly, let us find the generalized inverse temperatures $\beta_\kappa$ in the GGE for harmonic systems. These are determined from Eq.~\eqref{betakappa}. Since all the charges $h_\kappa$ live in non-overlapping Hilbert spaces, we have
\bea \nonumber
	\tr\left[h_\kappa \rho_{AB}^\GGE \right] = \frac{\tr\left[h_\kappa e^{-\beta_\kappa h_\kappa}\right]}{\tr\left[ e^{-\beta_\kappa h_\kappa} \right]} = \frac{\Omega_\kappa}{2}
\coth \frac{\beta_\kappa \Omega_\kappa}{2}.
\eea
Thus, equating this to $\av{h_\kappa} = \tr[h_\kappa \rho_{AB}(0)])$, we find
\bea \label{ugly}
	\beta_\kappa = \frac{2}{\Omega_\kappa} \arctanh \frac{\Omega_k}{2 \av{h_\kappa}}.
\eea

\section{G-LOCAL THERMALITY FOR EDGE--EDGE COUPLED LATTICES}
\label{app:EEdyn}

EE coupling is present when the interaction Hamiltonian is of the form
\bea \label{hint_EE}
	H_\I^{\mathrm{(EE)}} = \lambda \sum_{\nu \in \mathrm{edge}} q_{A, \nu} q_{B, \nu} ,
\eea
where $\nu$ runs over the sites located on the interacting edge of each lattice (see left column of Fig.~\ref{fig:couplings} for an illustration).

For example, when $A$ and $B$ are 1D, the edge consists of a single site. For such a configuration with $A$ and $B$ featuring nearest-neighbor interactions, the dynamics of the fidelity $\cF_s^{\max}$ and effective {\glocal} temperature $T_s^\eff$ of a single-site subsystem $s$ located at the center of $X$ is plotted in Fig.~\ref{fig:gloc_time_1DEE}. Similarly to the case of FB coupling discussed in Sec.~\ref{sec:glocality}, we see that $s$ remains {\glocal}ly thermal at all times with a good approximation.

We also notice on the bottom panel of Fig.~\ref{fig:gloc_time_1DEE} that the effective temperature of $s$ remains unchanged for some time. This happens because the speed of sound in each system is finite, and therefore it takes a finite amount of time until the perturbation caused by switching on the coupling at the edge to reach the center of the chain (where $s$ is).

For this very reason, there is also a temperature gradient within each lattice $X$ in the transient regime, before the total system equilibrates. A snapshot of that is presented in Fig.~\ref{fig:gloc_site_1DEE}, where the fidelity $\cF_s^{\max}$ and effective {\glocal} temperature $T_s^\eff$ are plotted as a function of the position of a single-site subsystem $s$ that slides along the chain (just like in Fig.~\ref{fig:gloc_site_1DFB}). Here we see that, while all $s$ are {\glocal}ly thermal with excellent approximation, their temperature now depends on the position of $s$. Due to this gradient, the effective canonical temperature \eqref{Teff_can} becomes inadequate, as is emphasized in the inset of the bottom panel of Fig.~\ref{fig:gloc_time_1DEE}.

Expectedly, at those times when there is a temperature gradient in the lattices, the decay of $\cF_s^{\max}$ with $N_s$ is faster as compared with the FB case. Moreover, even small (e.g., $N_s = 2$) but delocalized subsystems $s$ (i.e., when $s = \{\nu_1, \nu_2\}$ with, e.g., $\nu_1 = 50$ and $\nu_2 = 150$), are not {\glocal}ly thermal anymore. This contrasts the [1D, FB] and [2D, FB] cases, where all small subsystems, localized or not, are {\glocal}ly thermal with good approximation.

\section{NUMERICAL DEMONSTRATION OF THE MAIN RESULT}
\label{app:numerics}

In this section, we numerically demonstrate the validity of our main result laid out in Sec.~\ref{sec:glocality} (and illustrated in Fig.~\ref{fig:sketch}). It states that {\glocal} thermality of $A$ and $B$ is guaranteed \textit{at all times, including during the transient}, whenever all local observables of $AB$ equilibrate dynamically at long times.

We will first describe the parameter space and then discuss the relevant figures of merit and show pertinent results of our simulations.

\subsection{Parameterization}
\label{app:param}

A natural dimensionless parametrization of the system and its dynamics can be achieved as follows. First of all, we recall that we work in the natural units where $\hbar = k_{\mathrm{B}} = 1$ and the masses of all the oscillators are set to $1$. Therefore, the transformation $\tilde{q} = q \sqrt{\omega}$, $\tilde{p} = p / \sqrt{\omega}$ will render $\tilde{q}$ and $\tilde{p}$ dimensionless while preserving the canonical commutation relations. In these terms,
\bea
H_X = \omega_X \, h_X(\tilde{q}_\nu, \tilde{p}_\nu, \widetilde{g}_X, \alpha),
\eea
where the dimensionless operator function $h_X$ of the dimensionless quantities $(\tilde{q}_\nu, \tilde{p}_\nu, \widetilde{g}_X, \alpha)$,
with
\bea
\widetilde{g}_X := g_X / \omega_X^2,
\eea
is given by
\bea
h_X = \frac{1}{2} \sum_\nu \big( \tilde{q}_\nu^2 + \tilde{p}_\nu^2 \big) + \sum_{\nu \neq \nu'} \widetilde{G}_X^{\nu, \nu'} \tilde{q}_{X, \nu} \tilde{q}_{X, \nu'}.
\eea
Here, by natural extension of Eq.~\eqref{decayrate},
\bea
\widetilde{G}_X^{\nu, \nu'} := \frac{\widetilde{g}_X}{\dist(\nu, \nu')^\alpha}.
\eea

Introducing the dimensionless lattice--lattice coupling
\bea
\widetilde{\lambda} := \lambda / (\omega_A \omega_B),
\eea
and
\bea
\mu := \sqrt{\omega_A / \omega_B},
\eea
we obtain the total Hamiltonian 
\bea
H_\tot = \sqrt{\omega_A \omega_B} \, h_\tot,
\eea
where, for e.g. the FB coupling, the dimensionless operator $h_\tot$ is
\bea
h_\tot = \mu h_A + \mu^{-1} h_B + \widetilde{\lambda} \sum_\nu \tilde{q}_{A, \nu} \tilde{q}_{B, \nu} .
\eea
As mentioned in Sec.~\ref{sec:setup}, in the $5$-dimensional system-parameter space with coordinates $(\mu, \alpha, \widetilde{\lambda}, \widetilde{g}_A, \widetilde{g}_B)$, the set of allowed system parameters is determined by the condition that the operator $h_\tot$ is unbounded from below.

Lastly, the evolution in dimensionless time
\bea
\tilde{t} := t \sqrt{\omega_A \omega_B}
\eea
is generated by
\bea
U = e^{- i \tilde{t} h_\tot}.
\eea
And defining the dimensionless temperatures as
\bea
\widetilde{T}_X := T_X / \omega_X,
\eea
we can express the initial state \eqref{rhonin} in terms of dimensionless quantities:
\bea
\rho_{AB}(0) \propto e^{- h_A / \widetilde{T}_A} \otimes e^{- h_B / \widetilde{T}_B}.
\eea

\subsection{Relevant quantities and data}
\label{app:quantifex}

Although we formulated our all-time g-local thermality result in a ``discrete'' true--false language (see Fig.~\ref{fig:sketch}), there is more quantitative structure to the dependence of the degree of g-local thermality on the degree of long-time equilibration. To properly showcase this relationship, we need a quantification of both phenomena.

First of all, we pick a long enough time interval $[0, \TT_m]$ over which we observe the system. Then, since we already have a well-defined measure of g-local thermality at an instant of time $\TT$ and subsystem $s$, $\cF^{\max}_s(\TT \,)$, we use it to introduce a measure of all-time (AT) g-local thermality at $s$ defined as
\bea
\cF^\AT_s = \min_{\TT \in [0, \, \TT_m]} \cF^{\max}_s(\TT \,).
\eea

To quantify the degree to which the system locally equilibrates as per the definition in Sec.~\ref{sec:equil}, we will employ the fact that, if equilibration occurs, then it is described by the GGE in the thermodynamic limit (see Sec.~\ref{sec:equil}). The main quantifier here is the longest ``equilibrium interval'' during the $[0, \TT_m]$ period. By an equilibrium interval we mean any $[\TT^{\,\,\equi}_{\mathrm{i}}, \TT^{\,\,\equi}_{\mathrm{f}}]$ such that $\cD[\rho_s(\TT\,), \rho_s^\GGE] \leq \epsilon$ $\forall \TT \in [\TT^{\,\,\equi}_{\mathrm{i}}, \TT^{\,\,\equi}_{\mathrm{f}}]$. The figure of merit we will use is the ratio of the longest equilibrium interval,
\bea
\tau_s^\equi := \max_{[\TT^{\,\,\equi}_{\mathrm{i}}, \, \TT^{\,\,\equi}_{\mathrm{f}}] \subset [0, \, \TT_m] } \; (\TT^{\,\,\equi}_{\mathrm{f}} - \TT^{\,\,\equi}_{\mathrm{i}})
\eea
to the total duration of observation:
\bea
r^\equi_s = \frac{\tau_s^\equi}{\TT_m}.
\eea

In parallel with $r_s^\equi$, we will use the average distance from equilibrium,
\bea
\av{\cD_s} = \frac{1}{\TT_m} \int_0^{\TT_m} d\TT \, \cD[\rho_s(\TT \,), \rho_s^\GGE],
\eea
to quantify local equilibration.

The quantity $\langle \cD_s \rangle$ cannot be used alone to ``measure'' equilibration, as even a very small value of $\langle \cD_s \rangle$ does not exclude frequent $\epsilon$-surpassing peaks of $\cD[\rho_s(\TT \,), \rho_s^\GGE]$. Similarly, used alone, $r^\equi_s$ indicates the time $\cD_s$ uninterruptedly spends under $\epsilon$, but does not tell us how much lower than $\epsilon$ it typically gets. So, although $r^\equi_s$ and $\av{\cD_s}$ are not independent (e.g., if $r_s^\equi = 1$, then necessarily $\av{\cD_s} \leq \epsilon$), only when considering them together does one get a complete picture of how well $AB$ locally equilibrates at $s$---one needs a small $\langle \cD_s \rangle$ and a large $r^\equi_s$ to ensure equilibration.

\begin{figure}[!t]
\centering
\includegraphics[width = 0.98\columnwidth]{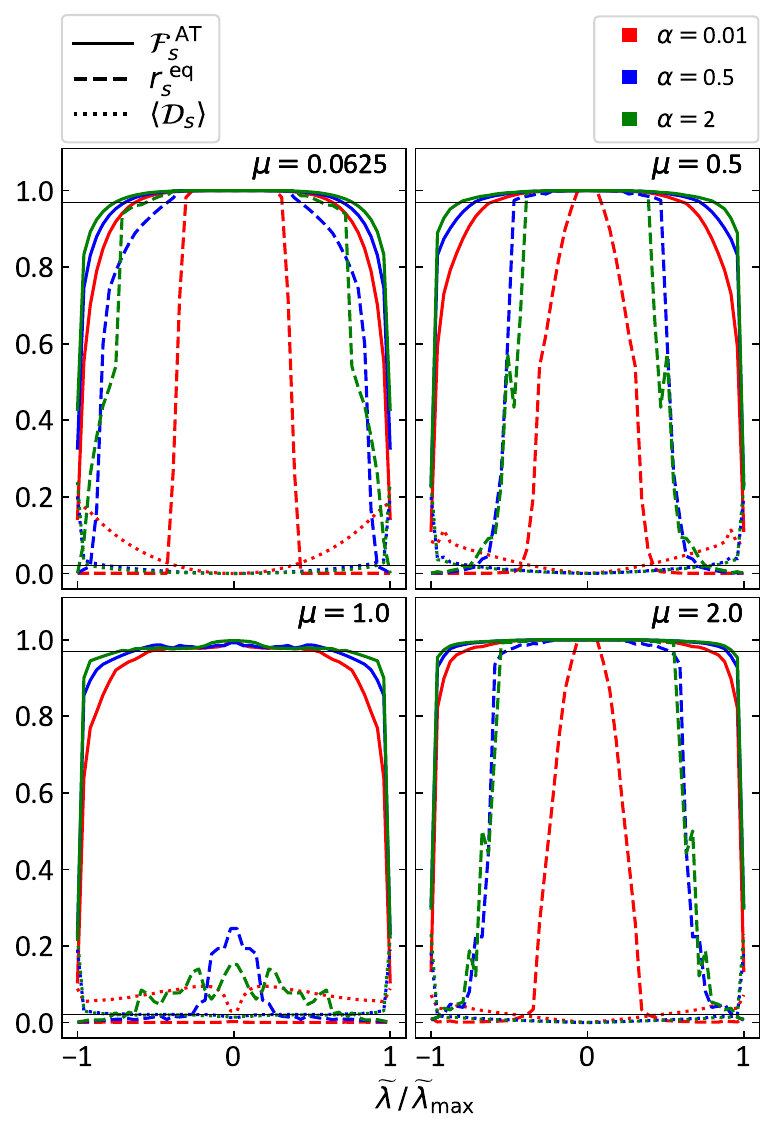}
\caption{
(\textbf{Degrees of all-time {\glocal} thermality and long-time equilibration vs $\widetilde{\lambda}$.)} Each panel shows $\cF_s^\AT$ (solid line), $r_s^\equi$ (dashed line), and $\av{\cD_s}$ (dotted line) as functions of $\widetilde{\lambda}$. Different colors correspond to different values of $\alpha$. The upper horizontal line is at $0.97$ and represents the threshold value of $\cF_s^\AT$ above which we say that the system is g-locally thermal at all times. The lower horizontal line is the $\epsilon = 0.02$ threshold for $\av{\cD_s}$. The configuration is [1D, FB], and the plots are for the central two-node subsystem of $A$. All the other parameters are $N_A = N_B = 200$, $\widetilde{g}_A = 0.2$, $\widetilde{g}_A = 0.3$, $\widetilde{T}_A = 0.1$, $\widetilde{T}_B = 1$. The $\mu = 1$ ($\omega_A = \omega_B$) case is peculiar in that the equal frequencies result in resonant oscillations that give rise to periodic spikes in $\cD[\rho_s(t), \rho_s^\GGE]$, resulting in poor equilibration. The system experiences all-time g-local thermality to a good extent nonetheless. Viewing the plot from $\widetilde{\lambda} = 0$, we see that the quality of equilibration deteriorates much faster than the degree of all-time g-local thermality as $\widetilde{\lambda}$ approaches $\pm \widetilde{\lambda}_{\max}$---the maximal value of $|\widetilde{\lambda}|$ for which $H_\tot$ is bounded from below.
}
\label{fig:GLT_mula}
\end{figure}

Regarding the choice of $\TT_m$, we note that, although verifying equilibration is in general an undecidable problem \cite{Shiraishi_2021}, the situation in quadratic harmonic systems is more predictable. Indeed, as discussed in Sec.~\ref{sec:equil}, if equilibration occurs, then the equilibrium is described by the GGE. Moreover, it was shown in Ref.~\cite{Gluza_2019} that, if the interactions in the system are of short range, then the equilibration time does not depend on the system size and the recurrence time grows linearly with the size. For systems with long-range interactions, our numerical experiments show that the pattern is similar---if the system equilibrates, it does so relatively quickly; and if it does not, then local states show no tendency to converge at long times. For the plots in Figs.~\ref{fig:GLT_mula}-\ref{fig:TRUE-FALSE_GLT_gAgB}, we found $\TT_m = 10000$ to be sufficiently long, yet not too long for recurrences to significantly affect the picture.

\definecolor{mycyan}{rgb}{0,1,1}
\definecolor{mymagenta}{rgb}{1,0,1}
\definecolor{mypurple}{rgb}{0.727,0.125,0.741}
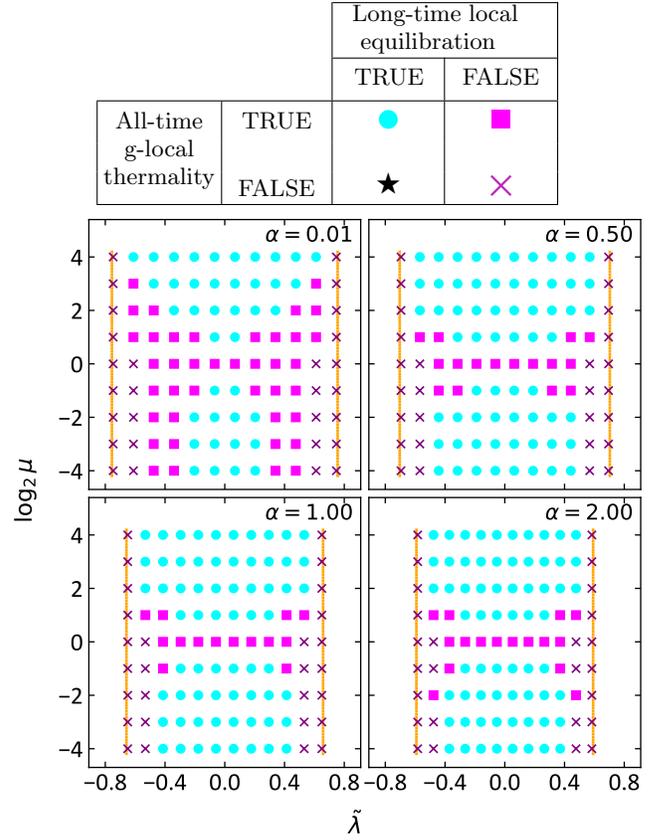
\begin{figure}[t!]
\centering
\begin{subfigure}{\columnwidth}
\centering
\begin{tabular}{|p{1.5cm}|c|c|c|}
 \cline{3-4}
 \multicolumn{2}{c|}{} & \multicolumn{2}{p{2.4cm}|}{\centering \; Long-time local equilibration}
 \\
 \cline{3-4}
 \multicolumn{2}{c|}{} & \multicolumn{1}{c|}{\,\, TRUE $\phantom{\Big\vert}$} & \multicolumn{1}{c|}{\, FALSE \,}
 \\
 \hline
 \multirow{2}{1.5cm}{\centering All-time g-local thermality} & $\phantom{\Big\vert}$ TRUE $\phantom{\Big\vert}$ & \begin{tikzpicture}    
        \fill[mycyan] (0,0) circle[radius=0.125];
    \end{tikzpicture} & \begin{tikzpicture}   
        \fill[mymagenta] (0,0) rectangle (0.25,0.25);
    \end{tikzpicture}
    \\
    & \multicolumn{1}{c|}{} & \multicolumn{1}{c|}{} & \multicolumn{1}{c|}{}
    \\
 & FALSE & \FiveStar & \textcolor{mypurple}{\footnotesize{\XSolid}}
 \\
 \hline
\end{tabular}
\end{subfigure}
\hfill
\begin{subfigure}{\columnwidth}
\includegraphics[width=\columnwidth]{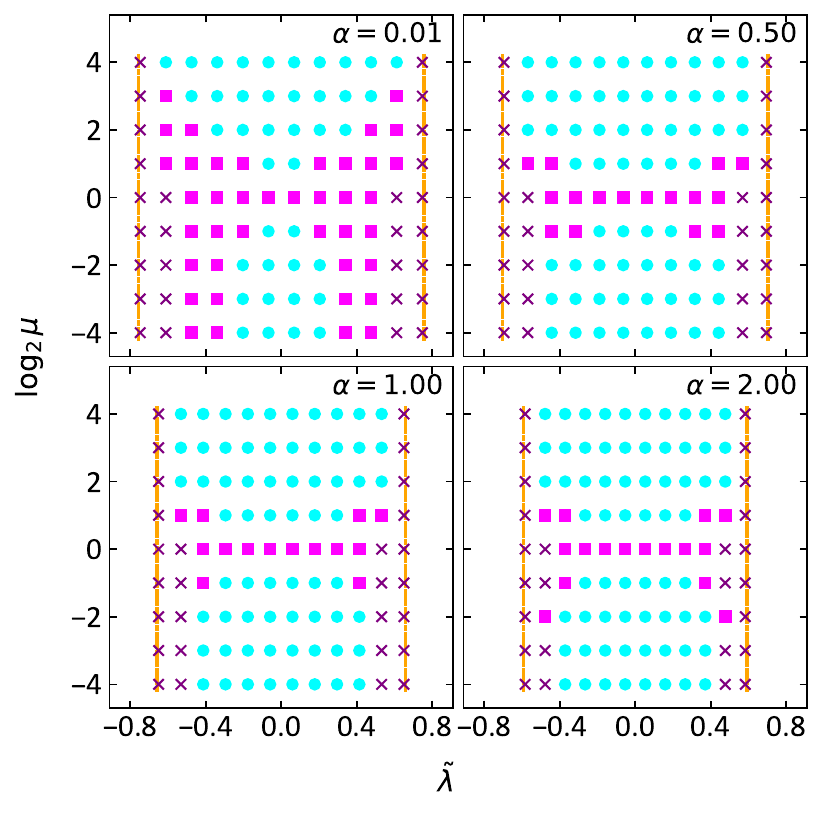}
\end{subfigure}
    \caption{
    (\textbf{All-time {\glocal} thermality and long-time local equilibration vs $\widetilde{\lambda}$ and $\mu$}.) Each panel shows whether or not all-time {\glocal} thermality and long-time equilibration occur, as per the color- and shape-coding in the table, as a function of $\widetilde{\lambda}$ and $\mu$. The panels are for four different values of $\alpha$, with all other parameters fixed to the same values as in Fig.~\ref{fig:GLT_mula}. The orange lines are at $\pm \widetilde{\lambda}_{\max}$, marking the boundary of the set of all $(\widetilde{\lambda}, \, \mu)$ for which $H_\tot$ is bounded from below. Here the TRUE-threshold for all-time {\glocal} thermality is chosen to be $\cF_s^\AT \geq 0.97$ for all subsystems $s$ of size two. For long-time equilibration, the TRUE-threshold is $\min\limits_{|s|=2} r_s^\equi \geq 0.8$ \textbf{and} $\max\limits_{|s|=2} \av{\cD_s} \leq \epsilon = 0.02$. We see that all combinations occur except the {\FiveStar}, in accordance with the claim in the main text; see also Fig.~\ref{fig:sketch}.
    \textbf{The table} shows the shape- and color-coding of the four logical possibilities for observing or not observing all-time {\glocal} thermality and long-time local equilibration used in the plot.
    }
\label{fig:TRUE-FALSE_GLT_mula}
    \end{figure}


\begin{figure}[!t]
\centering
\includegraphics[width = \columnwidth]{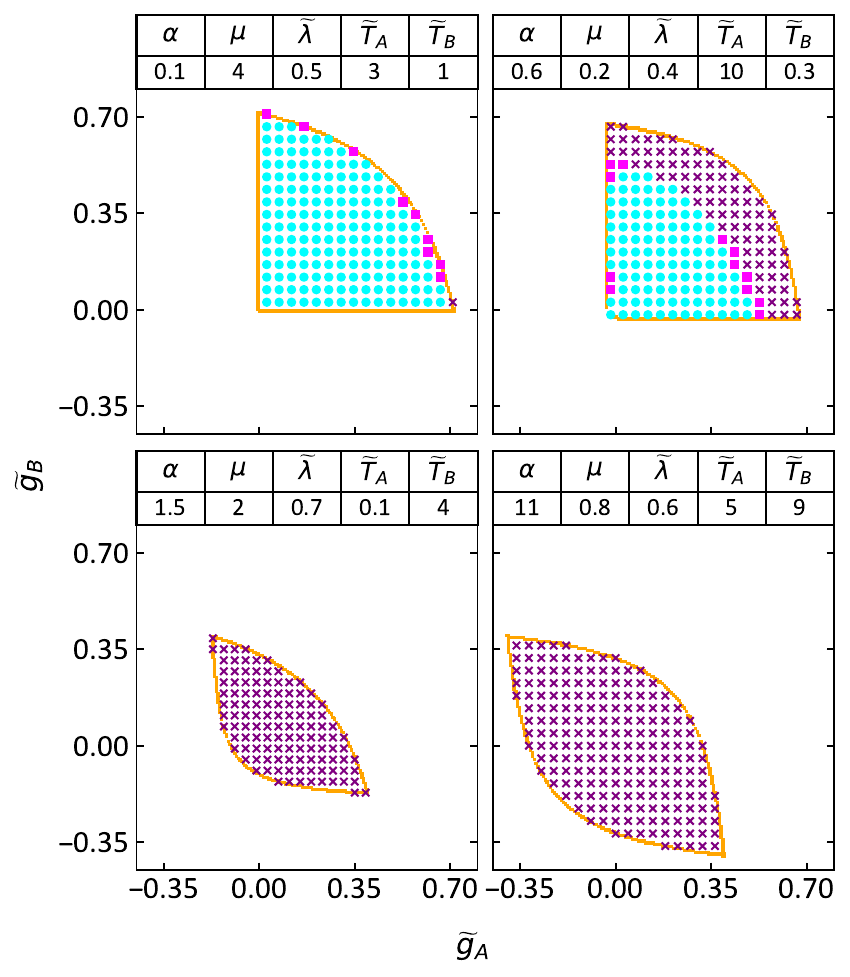}
\caption{
(\textbf{All-time {\glocal} thermality and long-time local equilibration vs $\widetilde{g}_A$ and $\widetilde{g}_B$.)} Each panel shows whether or not all-time {\glocal} thermality and long-time equilibration occur, as per the color- and shape-coding in the table in Fig.~\ref{fig:TRUE-FALSE_GLT_mula}, as a function of $\widetilde{g}_A$ and $\widetilde{g}_B$. The panels are for four different sets of parameters $(\alpha, \mu, \widetilde{\lambda}, \widetilde{T}_A, \widetilde{T}_B)$ shown on top of each panel. As previously, $N_A = N_B = 200$ and the configuration is [1D, FB]. On each panel, the orange border outlines the set of all $(\widetilde{g}_A, \widetilde{g}_B)$ such that $H_\tot$ is bounded from below. As in Fig.~\ref{fig:TRUE-FALSE_GLT_mula}, here the TRUE-threshold for all-time {\glocal} thermality is $\min\limits_{|s|=2} \cF_s^\AT \geq 0.97$, and for long-time equilibration, the TRUE-threshold is $\min\limits_{|s|=2} r_s^\equi \geq 0.8$ \textbf{and} $\max\limits_{|s|=2} \av{\cD_s} \leq \epsilon = 0.02$. We see that all combinations occur except the {\FiveStar}, in accordance with the claim illustrated in Fig.~\ref{fig:sketch}. The lack of all-time g-local thermality and equilibration in two bottom panels is consistent with Fig.~\ref{fig:GLT_mula}: in both panels, $\widetilde{\lambda}/ \widetilde{\lambda}_{\max}$ is close to $1$.
}
\label{fig:TRUE-FALSE_GLT_gAgB}
\end{figure}

In terms of the above-introduced quantities, the claim in our main result is as follows (see Fig.~\ref{fig:sketch} for a recap). If $r_s^\equi$ is small \textbf{or} $\av{\cD_s}$ is large (i.e., poor equilibration) then $\cF^\AT_s$ can be anything. If $r_s^\equi$ is close to $1$ \textbf{and} $\av{\cD_s}$ is small (i.e., good equilibration), then $\cF^\AT_s$ has to be close to $1$.

This is indeed what we see numerically. We have explored the full parameter range by both randomly sampling the parameters $(\mu, \alpha, \widetilde{\lambda}, \widetilde{g}_A, \widetilde{g}_B, \widetilde{T}_A, \widetilde{T}_B)$ and deliberately choosing values at the boundaries of the set of allowed parameters (determined by the condition that the total Hamiltonian is nonnegative; see Appendix~\ref{app:param}).

Since the parameter space is $7$-dimensional (not counting $N_A$, $N_B$, and the configuration of lattice--lattice interaction), and therefore impossible to draw, we will present our results in two-dimensional cross-sections.

In our numerical experiments, we found that there are three ``dangerous'' parameter regimes. First is strong $A/B$ frequency imbalance: $\mu \ll 1$ or $\mu \gg 1$. The other parameter that has a significant effect on g-local-thermality--equilibration relation is $\alpha$---the range of intra-lattice interactions. Indeed, as we will see in Appendix~\ref{app:brandrammer}, canonical-typicality--based results become inapplicable for small values of $\alpha$, making the regime of small $\alpha$'s also dangerous. The third is the regime of strong lattice--lattice coupling, which also bears the potential to be dangerous as both $A$ and $B$ lose their dynamic individuality when $\widetilde{\lambda}$ is large, especially in the FB coupling configuration.
Therefore, our emphasis will be on cross-sections of $\mu$, $\alpha$, and $\widetilde{\lambda}$.
Note that large $\widetilde{\lambda}$ means that $|\widetilde{\lambda}|$ is close to $\widetilde{\lambda}_{\max}$, which is the maximal value of $|\widetilde{\lambda}|$ (with all other parameters fixed) for which $H_\tot$ is bounded from below.

The most insight is provided by Fig.~\ref{fig:GLT_mula}, where we plot the degrees of all-time g-local thermality ($\cF_s^\AT$) and equilibration ($r_s^\equi$ and $\av{\cD_s}$) as functions of $\widetilde{\lambda}$, for different (extremal and not) values of $\alpha$ and $\mu$. There we see that the quality of equilibration deteriorates significantly faster than the degree of all-time g-local thermality as $|\widetilde{\lambda}|$ approaches its maximum ($\widetilde{\lambda}_{\max}$). This confirms our claim and, in a way, makes the relation between all-time g-local thermality and long-time equilibration more quantitative.

A different cross-section of the $(\mu, \alpha, \widetilde{\lambda})$ subset is presented in Fig.~\ref{fig:TRUE-FALSE_GLT_mula}. There we plot all-time g-local thermality and long-time local equilibration in a discrete fashion: occurs (TRUE) or does not occur (FALSE). The four logical possibilities are presented in the table in Fig.~\ref{fig:TRUE-FALSE_GLT_mula}, with corresponding color- and shape-coding. Now, the claim of our main result, as formulated in Sec.~\ref{sec:glocality} and summarized in Fig.~\ref{fig:sketch}, is that the combination [all-time g-local thermality = FALSE] \textbf{and} [long-time local equilibration = TRUE], encoded as {\FiveStar}, never occurs. And indeed, we see no {\FiveStar} in Fig.~\ref{fig:TRUE-FALSE_GLT_mula}.

Figure~\ref{fig:TRUE-FALSE_GLT_gAgB} is another TRUE--FALSE plot, this time across four $(\widetilde{g}_A, \widetilde{g}_B)$ planes, each corresponding to a panel of the plot. The values of the parameters fixing the planes are presented above the panels. In the top left panel, we have $A$ with much larger initial energy and heat capacity ($\omega_A = 16 \, \omega_B$ and $T_A = 48 \, T_B$), so, expectedly, the evolution does not perturb it much, so its g-local thermality is largely maintained even at the edge of the parameter space. On the top right panel, $A$ has a much smaller heat capacity than $B$ ($\omega_A = \omega_B/25$) and starts at a similar temperature with $B$ ($T_A \approx 1.33 T_B$), so we see more diversity of options. In two bottom panels, the respective $\widetilde{\lambda}$'s are close to their maximal values, so, as could be anticipated from Fig.~\ref{fig:GLT_mula}, we observe that neither all-time g-local thermality nor long-time local equilibration occur to a high enough degree.

We observe the picture described above over all cross-sections of the parameter set, for all configurations and spatial dimensions. We take this as a compelling numerical proof of our main result.

\section{COMPARISON WITH ENSEMBLE EQUIVALENCE}
\label{app:brandrammer}

By the stronger equivalence of ensembles, we mean Proposition 2 in Ref.~\cite{Brandao_2015}. In a slightly simplified form derived in Ref.~\cite{Farrelly_2017} (Lemma 2), it states the following. Say, $X$ is a $d$-dimensional (hyper)cubic lattice with $N = n^d$ sites. Each site contains a quantum system described by a finite-dimensional Hilbert space, with the dimension being the same for all sites. The Hamiltonian is of finite range (i.e., local, as per the definition in Appendix~\ref{app:local_meas}): $H_X = \sum_{\nu \in X} \mathfrak{h}_\nu$, with $\mathfrak{h}_\nu$ acting only on sites $\nu'$ with $\dist(\nu, \nu') \leq k$. Now, let $\tau_X$ be a state with exponentially decaying correlations and fix some $0< c < 1/(d+2)$. If
\bea \label{relent}
	S(\rho_X \Vert \tau_X) = o \big( N^{\frac{1- c (d+2)}{d+1}} \big),
\eea
where $S(\rho \Vert \tau) :=\tr[\rho (\ln \rho - \ln \tau)]$ is the relative entropy, then \cite{Brandao_2015, Farrelly_2017}
\bea \label{BC_main}
	\mathbbm{E}_{s \in C_l} \, \Vert \rho_{s} - \tau_{s} \Vert_1 = O\big( N^{-c/2} \big),
\eea
where $\Vert \mathfrak{O} \Vert_1 := \tr \sqrt{\mathfrak{O}^\T \mathfrak{O}}$ is the trace norm, $C_l$ is the set of all sub-hypercubes $s$ of $X$ with side length
\bea \label{BC_subsize}
	l = o\big(n^{\frac{1-c}{d+1}} \big),
\eea
and $\mathbbm{E}_{s \in C_l}$ denotes arithmetic averaging over $C_l$. Namely,
\bea
	\mathbbm{E}_{s \in C_l} \, \Vert \rho_{s} - \tau_{s} \Vert_1 := \frac{1}{K_l} \sum_{s \in C_l} \Vert \rho_{s} - \tau_{s} \Vert_1,
\eea
where $K_l$ is the size of $C_l$ (i.e., the amount of sub-hypercubes $s$ in it). Here, the big-$O$ and small-$o$ are as per the standard asymptotic notation, and, as in the main text, $\rho_{s} := \tr_{X \backslash s} [\rho_X]$.

In simple terms, this lemma means that, if $\tau_X$ has exponentially decaying correlations, and $\rho_X$ is not very far from it in terms of the relative entropy [Eq.~\eqref{relent}], then $\rho_X$ is locally close to $\tau_X$, in trace norm, for almost all small subsystems [Eq.~\eqref{BC_main}]. Here, small is any subsystem the diameter of which is $o\big(n^{\frac{1-c}{d+1}} \big)$ [Eq.~\eqref{BC_subsize}]; of course, any fixed size is $o\big(n^{\frac{1-c}{d+1}} \big)$ in the thermodynamic limit.

To translate Eq.~\eqref{BC_main} into a statement about the Bures distance, note that, in view of the Fuchs--van de Graaf inequality \cite{Fuchs_1999}, $\cD[\rho, \tau]^2 \leq \Vert \rho - \tau\Vert_1$. Therefore,
\beaa \nonumber
	\mathbbm{E}_{s \in C_l} \, \cD[\rho_s, \tau_s] &= \frac{1}{K_l} \sum_{s \in C_l} 1 \cdot \cD[\rho_s, \tau_s]
\\
&\stackrel{(*)}{\leq} \frac{1}{K_l} \sqrt{\sum_{s \in C_l} 1^2} \sqrt{\sum_{s \in C_l} \cD[\rho_s, \tau_s]^2}
\\
&\leq \sqrt{\frac{1}{K_l} \sum_{s \in C_l} \Vert \rho_s - \tau_s \Vert_1},
\eeaa
where the step $(*)$ is due to the Cauchy--Schwarz inequality. Hence, in view of Eq.~\eqref{BC_main}, we find that
\bea \label{BC_adapted}
	\mathbbm{E}_{s \in C_l} \, \cD[\rho_s, \tau_s] = O\big( N^{-c/4} \big).
\eea

Coming back to our setup, let $\rho_X(t)$ be the state of the lattice $X$ at the moment of time $t$, and $T_X^{\eff, \can}(t)$ be its effective canonical temperature, as per the definition \eqref{Teff_can}. Also let
\bea
	\tau_X^{\eff, \can}(t) := \tau\big( T_X^{\eff, \can}(t), H_X \big)
\eea
be the Gibbs state corresponding to it. Now, observing that, due to Eq.~\eqref{Teff_can},
\bea \nonumber
	S\big( \rho_X(t) \, \big\Vert \, \tau_X^{\eff, \can}(t) \big) = S\big( \tau_X^{\eff, \can}(t) \big) - S(\rho_X(t)),
\eea
where $S(\rho)$ is the von Neumman entropy, and keeping in mind Eqs.~\eqref{BC_adapted} and~\eqref{def:g-local}, we can state the following consequence of Proposition 2 of Ref.~\cite{Brandao_2015} and Lemma 2 of Ref.~\cite{Farrelly_2017}.
\begin{corollary}[of Proposition 2 of Ref.~\cite{Brandao_2015}]
\label{thm:BC_corollary}
If $\tau_X^{\eff, \can}(t)$ has exponentially decaying correlations and
\bea \label{BC_adapted_cond}
	S\big( \tau_X^{\eff, \can}(t) \big) - S(\rho_X(t)) = o \big( N^{\frac{1- c (d+2)}{d+1}} \big), ~~
\eea
then $\rho_X(t)$ is {\glocal}ly thermal at (almost) uniform temperature $T_X^{\eff, \can}(t)$, up to a correction $\propto N^{-c/4}$. The {\glocal} thermality of $X$ is at the level of subsystems of diameter $l = o\big(n^{\frac{1-c}{d+1}} \big)$ [Eq.~\eqref{BC_subsize}].
\end{corollary}

This result provides a background against which we can assess how ``expected'' the all-time {\glocal} thermality result in Sec.~\ref{sec:glocality} is for short-range Hamiltonians. Indeed, when $H_X$ is finite-ranged and gapped, $\tau(T, H_X)$ has exponentially decaying correlations at any $T$ \cite{Cramer_2006}. So, the first condition of Corollary~\ref{thm:BC_corollary} is satisfied.

The validity of the second condition is in general much harder to assess a priori. But given that the entropy difference in Eq.~\eqref{BC_adapted_cond} is zero at $t=0$ in our setting, it would not be too surprising if it were to remain small enough to never violate Eq.~\eqref{BC_adapted_cond} during the evolution. This is so especially when $H_\I$ is invariant under translations in $X$ (e.g., FB coupling), since in that case, one expects that both $\tau_X^{\eff, \can}(t)$ and $\rho_X(t)$ will remain translationally invariant (except for the edges) at all times.

We emphasize that Corollary~\ref{thm:BC_corollary} has no bearing on the degree of validity of the condition \eqref{BC_adapted_cond}. Thus, it does \textit{not} prove all-time {\glocal} thermality of $X$ when $H_X$ is short-ranged and gapped and $H_\I$ is translationally invariant. However, the Corollary does show that it is not unexpected that we do observe all-time {\glocal} thermality for such systems. In fact, when $A$ and $B$ are FB-coupled 2D lattices with nearest-neighbor interactions, we see in the bottom inset of Fig.~\ref{fig:gloc_time_2DFB} that {\glocal} thermality is accompanied by $T_X^\eff(t) \approx T_X^{\eff, \can}$ at all times. This suggests (but does not prove) that Corollary~\ref{thm:BC_corollary} applies in this case.

The situation changes, even for short-ranged $H_X$'s, when $H_\I$ is not translationally invariant; e.g., in the case of EE coupling. Then, due to the gradients of energy (and temperature), the condition \eqref{BC_adapted_cond} becomes unlikely to be satisfied at all times. By directly looking at $\mathbbm{E}_{s \in C_l} \, \cD[\rho_s, \tau_s]$, we indeed see that Eq.~\eqref{BC_adapted} is violated. Let us consider, for concreteness, the [1D, EE] case for which Fig.~\ref{fig:gloc_site_1DEE} is plotted. Each single-site subsystem is almost exactly {\glocal}ly thermal, so, noticing that $C_1$ is simply the set of all sites of $X$, we can write
\bea \nonumber
	\mathbbm{E}_{s \in C_1} \, \cD[\rho_s, \tau_s] = \frac{1}{N} \sum_{\nu=1}^N \cD\big[ \tau_\nu^\MF\big(T^\eff_\nu\big), \tau_\nu^\MF\big(T_X^{\eff, \can}\big) \big].
\eea
And since almost half of the sites of $A$ are at $T_\nu^\eff \approx 0.5$ and the other $\approx 25\%$ are at $T_\nu^\eff \approx 0.1$, there exists some $\zeta > 0$ such that $\sum_{\nu=1}^N \cD\big[ \tau_\nu^\MF\big(T^\eff_\nu\big), \tau_\nu^\MF\big(T_X^{\eff, \can}\big) \big] \geq N \zeta$. Hence, $\mathbbm{E}_{s \in C_1} \, \cD[\rho_s, \tau_s] \geq \zeta$, and therefore Eq.~\eqref{BC_adapted} cannot hold for $N \gg 1$. Thus, Corollary~\ref{thm:BC_corollary} does not apply. Nonetheless, both $A$ and $B$ remain {\glocal}ly thermal at all times, with very high accuracy, as Figs.~\ref{fig:gloc_time_1DEE} and \ref{fig:gloc_site_1DEE} clearly illustrate.

Even further from the scope of Ref.~\cite{Brandao_2015}, and consequently of Corollary~\ref{thm:BC_corollary}, are lattices with long-range interactions. In such systems, ensemble equivalence is known to generally fail \cite{Barre_2001, Campa_2009}. It is however worth noting that, in Ref.~\cite{Kuwahara_2020}, it is proven that, for long-range interacting systems, microcanonical and canonical ensembles become equivalent at high temperatures. However, i) that result is only for microcanonical and canonical states, not for general states like Proposition 2 of Ref.~\cite{Brandao_2015}, so it cannot be used in our scenario. And ii) for the values of interaction range $\alpha \leq d$, the threshold temperature above which the ensemble equivalence is established in Ref.~\cite{Kuwahara_2020} diverges with $N$. In contrast, our all-time {\glocal} thermality result holds for long-range interacting systems with arbitrary $\alpha$ (Figs.~\ref{fig:gloc_time_1DFB} and \ref{fig:gloc_site_1DFB} illustrate that for $\alpha = 0.5$).

That the ensemble equivalence fails, whereas all-time {\glocal} thermality persists, in the above two situations is a clear indication that the two phenomena are fundamentally different and independent from one another.

Lastly, we note that the results of Ref.~\cite{Brandao_2015}, and therefore Corollary~\ref{thm:BC_corollary}, are proven only for lattices of systems with finite Hilbert-space dimension. In our case, however, the on-site Hilbert-space dimension is infinite. This is not a problem at finite temperatures---the on-site oscillators can be approximated by finite systems by simply cutting off the nearly unpopulated high-energy states. However, the higher the temperatures, the higher the cut-off dimension has to be. And while the increase of the on-site Hilbert-space dimension significantly weakens the bounds established in Ref.~\cite{Brandao_2015}, interestingly, going to higher temperatures does not affect the precision of all-time {\glocal} thermality in all the systems we tested.

\section{RATE EQUATION IN TTM}
\label{app:monotonic}

According to the TTM \cite{Anisimov_1967, Anisimov_1974, Sanders_1977, Allen_1987, Jiang_2005, Lin_2008, Wang_2012, Liao_2014}, the energy exchange between two co-evolving systems is given by the equation
\beaa \label{rateeq1}
	\frac{dE_A}{dt} &= -(T_A - T_B) \, k(T_A, T_B),
\\
	\frac{dE_B}{dt} &= -(T_B - T_A) \, k(T_A, T_B),
\eeaa
where the thermal conductance $k(T_A, T_B)$ is positive provided heat flows from hot to cold.

Introducing the heat capacities for $A$ and $B$, respectively, as $C_A(T_A)$ and $C_B(T_B)$, we find the time evolution equation for the temperatures
(recalling that the TTM assumes that $A$ and $B$ are thermal at all times)
\beaa \label{rateeq2}
	\frac{dT_A}{dt} &= -(T_A - T_B) \frac{k(T_A, T_B)}{C_A(T_A)},
\\
	\frac{dT_B}{dt} &= -(T_B - T_A) \frac{k(T_A, T_B)}{C_B(T_B)}.
\eeaa
Keeping in mind that $C_A$ and $C_B$ are positive quantities, we introduce
\bea \nonumber
	J(T_A, T_B) := k(T_A, T_B) \Big[\frac{1}{C_A(T_A)} + \frac{1}{C_B(T_B)}\Big] \geq 0,
\\ \label{defJ}
\eea
and obtain
\bea
	\frac{d(T_A - T_B)}{dt} = - (T_A - T_B) J(T_A, T_B),
\eea
This leads us to
\bea \nonumber
	T_A(t) - T_B(t) = [T_A(0) - T_B(0)] \, e^{- \int_0^t ds \, J(T_A(s), T_B(s))},
\\ \label{monot1}
\eea
which indeed proves that $\left\vert T_A(t) - T_B(t) \right\vert$ monotonically decreases as the term in the exponent monotonically grows due to $J$ remaining positive at all times.

In fact, the reverse statement is also true: any two monotonically converging, differentiable functions $T_A(t)$ and $T_B(t)$ are a solution to a rate equation of the form \eqref{rateeq2}. To see this, we notice that one can ``reverse engineer'' Eq.~\eqref{monot1} to obtain the function $J(T_A(t), T_B(t))$, by taking the derivative of $T_A(t) - T_B(t)$. Next, since the heat capacities $C_A(T_A)$ and $C_B(T_B)$ are given, Eq.~\eqref{defJ} can be solved for $k(T_A(t), T_B(t))$. By construction, the solution of the equation \eqref{rateeq2} with the thus-obtained $k(T_A, T_B)$ will yield the given functions $T_A(t)$ and $T_B(t)$. Admittedly, this procedure is not useful in practical situations, it is presented here merely to prove the existence of a rate equation.

To summarize, the rate equation \eqref{rateeq1} and the monotonicity of the convergence of $T_A$ and $T_B$ are equivalent statements. Thus, the second postulate of the TTM can be equivalently rephrased as: ``the temperatures of the two systems approach each other monotonically in time.''

\section{ENERGY OF INTERACTION AT EQUILIBRIUM}
\label{app:energycondition}

\begin{figure}[!t]
\centering
\includegraphics[width = \columnwidth]{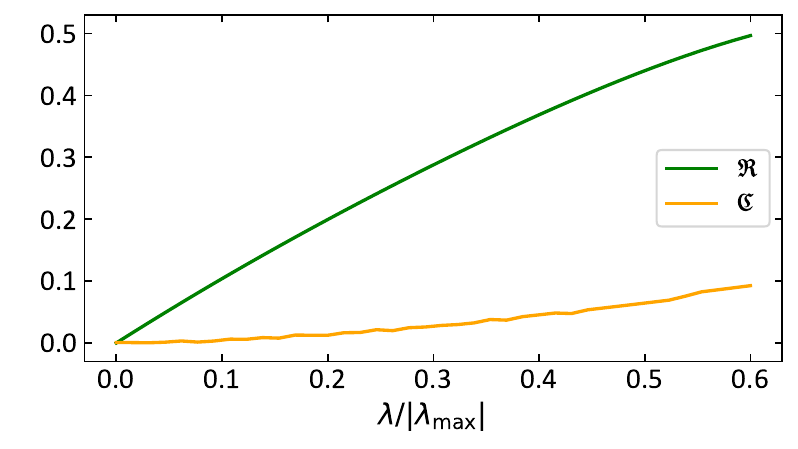}
\caption{
(\textbf{Role of interaction energy at equilibrium.})
The plot shows that while, at long times, $H_\I$ stores a significant amount of energy (green line), its effect on the validity of Eq.~\eqref{blimp} remains largely negligible even at strong couplings (orange line). Here $\mathfrak{R}$ [Eq.~\eqref{vayme}] measures the share of $H_\I$ in the energy balance at equilibrium and $\mathfrak{C}$ [Eq.~\eqref{oterim}] quantifies the precision to which Eq.~\eqref{blimp} is satisfied (with general $T_A^\equi$ and $T_B^\equi$). We see that $\mathfrak{C} \ll 1$ holds as long as the system equilibrates, and the choice of $\lambda \leq 0.6 \, |\lambda_{\max}|$ here ensures that it does. All other parameters are the same as in Fig.~\eqref{fig:heatflow}.
}
\label{fig:energybalance}
\end{figure}

Here we compare the share of $H_\I$ in the total energy at equilibrium with with the precision to which Eq.~\eqref{blimp} is satisfied.

The former is given by
\bea \label{vayme}
\mathfrak{R} := \frac{\av{H_\I}^\equi}{\av{H_A}^\equi + \av{H_B}^\equi},
\eea
where $\av{H_\I}^\equi = \tr[H_\I \rho^\GGE_{AB}]$, with the other two averages defined analogously. The ``precision'' of Eq.~\eqref{blimp} is quantified by
\bea \label{oterim}
\mathfrak{C} := 1 - \frac{\langle H_A\rangle_{T^{\eff, \equi}_A} + \langle H_B \rangle_{T^{\eff, \equi}_A}}{\langle H_A \rangle_{T_A} + \langle H_B \rangle_{T_B}},
\eea
with all the averages defined as in Eq.~\eqref{blimp}.

We numerically observe that $\mathfrak{C} \ll 1$ even for large values of $\lambda$, provided $AB$ equilibrates at long times. This is contrasted by the fact that $\av{H_\I}^\equi$ can be significant as compared to $\av{H_A}^\equi + \av{H_B}^\equi$. These aspects are illustrated in Fig.~\eqref{fig:energybalance}, where we see that $\mathfrak{C} \leq 0.1$ (in fact remaining much lower than $0.1$ up until $\lambda/\lambda_{\max} \approx 0.3$), whereas $\mathfrak{R}$ grows almost linearly with $\lambda$, becoming as large as $\approx 0.5$.

Lastly, note that, when $T^{\eff, \equi}_A \neq T^{\eff, \equi}_B$, Eq.~\eqref{blimp} will not be enough to determine both of these equilibrium temperatures.

\end{document}